\documentclass[fleqn,usenatbib]{mnras}

\usepackage{newtxtext,newtxmath}
\usepackage[dvipsnames]{xcolor}

\usepackage[T1]{fontenc}
\usepackage{subcaption}
\usepackage{ragged2e}

\DeclareRobustCommand{\VAN}[3]{#2}
\let\VANthebibliography\thebibliography
\def\thebibliography{\DeclareRobustCommand{\VAN}[3]{##3}\VANthebibliography}

\usepackage{graphicx}	
\usepackage{amsmath}	
\usepackage{soul}

\makeatletter
\let\jnl@style=\rm
\def\ref@jnl#1{{\jnl@style#1}}

\def\aj{\ref@jnl{AJ}}                   
\def\actaa{\ref@jnl{Acta Astron.}}      
\def\araa{\ref@jnl{ARA\&A}}             
\def\apj{\ref@jnl{ApJ}}                 
\def\apjl{\ref@jnl{ApJ}}                
\def\apjs{\ref@jnl{ApJS}}               
\def\ao{\ref@jnl{Appl.~Opt.}}           
\def\apss{\ref@jnl{Ap\&SS}}             
\def\aap{\ref@jnl{A\&A}}                
\def\aapr{\ref@jnl{A\&A~Rev.}}          
\def\aaps{\ref@jnl{A\&AS}}              
\def\azh{\ref@jnl{AZh}}                 
\def\baas{\ref@jnl{BAAS}}               
\def\bac{\ref@jnl{Bull. astr. Inst. Czechosl.}}
\def\caa{\ref@jnl{Chinese Astron. Astrophys.}}
\def\cjaa{\ref@jnl{Chinese J. Astron. Astrophys.}}
\def\icarus{\ref@jnl{Icarus}}           
\def\jcap{\ref@jnl{J. Cosmology Astropart. Phys.}}
\def\jrasc{\ref@jnl{JRASC}}             
\def\memras{\ref@jnl{MmRAS}}            
\def\mnras{\ref@jnl{MNRAS}}             
\def\na{\ref@jnl{New A}}                
\def\nar{\ref@jnl{New A Rev.}}          
\def\pra{\ref@jnl{Phys.~Rev.~A}}        
\def\prb{\ref@jnl{Phys.~Rev.~B}}        
\def\prc{\ref@jnl{Phys.~Rev.~C}}        
\def\prd{\ref@jnl{Phys.~Rev.~D}}        
\def\pre{\ref@jnl{Phys.~Rev.~E}}        
\def\prl{\ref@jnl{Phys.~Rev.~Lett.}}    
\def\pasa{\ref@jnl{PASA}}               
\def\pasp{\ref@jnl{PASP}}               
\def\pasj{\ref@jnl{PASJ}}               
\def\rmxaa{\ref@jnl{Rev. Mexicana Astron. Astrofis.}}%
\def\qjras{\ref@jnl{QJRAS}}             
\def\skytel{\ref@jnl{S\&T}}             
\def\solphys{\ref@jnl{Sol.~Phys.}}      
\def\sovast{\ref@jnl{Soviet~Ast.}}      
\def\ssr{\ref@jnl{Space~Sci.~Rev.}}     
\def\zap{\ref@jnl{ZAp}}                 
\def\nat{\ref@jnl{Nature}}              
\def\iaucirc{\ref@jnl{IAU~Circ.}}       
\def\aplett{\ref@jnl{Astrophys.~Lett.}} 
\def\apspr{\ref@jnl{Astrophys.~Space~Phys.~Res.}}
\def\bain{\ref@jnl{Bull.~Astron.~Inst.~Netherlands}} 
\def\fcp{\ref@jnl{Fund.~Cosmic~Phys.}}  
\def\gca{\ref@jnl{Geochim.~Cosmochim.~Acta}}   
\def\grl{\ref@jnl{Geophys.~Res.~Lett.}} 
\def\jcp{\ref@jnl{J.~Chem.~Phys.}}      
\def\jgr{\ref@jnl{J.~Geophys.~Res.}}    
\def\jqsrt{\ref@jnl{J.~Quant.~Spec.~Radiat.~Transf.}}
\def\memsai{\ref@jnl{Mem.~Soc.~Astron.~Italiana}}
\def\nphysa{\ref@jnl{Nucl.~Phys.~A}}   
\def\physrep{\ref@jnl{Phys.~Rep.}}   
\def\physscr{\ref@jnl{Phys.~Scr}}   
\def\planss{\ref@jnl{Planet.~Space~Sci.}}   
\def\procspie{\ref@jnl{Proc.~SPIE}}   

\makeatother

\title[Electromagnetic Counterparts of LISA EMRIs]{Repeating Nuclear Transients as Candidate Electromagnetic Counterparts of LISA Extreme Mass Ratio Inspirals}

\author[S. Kejriwal et al.]{
Shubham Kejriwal,$^{1}$\thanks{shubhamkejriwal@u.nus.edu (SK)}
Vojt\v{e}ch Witzany,$^{2}$\thanks{vojtech.witzany@matfyz.cuni.cz (VW)}
Michal Zaja\v{c}ek,$^{3}$\thanks{zajacek@physics.muni.cz (MZ)}
Dheeraj R. Pasham,$^{4}$\thanks{dheeraj@space.mit.edu  (DP)}
and Alvin J. K. Chua$^{1,5}$\thanks{alvincjk@nus.edu.sg (AC)}
\\
$^{1}$Department of Physics, National University of Singapore, Singapore 117551\\
$^{2}$Institute of Theoretical Physics, Faculty of Mathematics and Physics,
Charles University, CZ-180 00 Prague, Czech Republic\\
$^{3}$Department of Theoretical Physics and Astrophysics, Faculty of Science, Masaryk University, Kotl\'a\v{r}sk\'a 2, 611 37 Brno, Czech Republic\\
$^{4}$Kavli Institute for Astrophysics and Space Research, Massachusetts Institute of Technology, Cambridge, MA, USA\\
$^{5}$Department of Mathematics, National University of Singapore, Singapore 119076}

\pubyear{2024}

\begin{document}
\label{firstpage}
\pagerange{\pageref{firstpage}--\pageref{lastpage}}
\maketitle

\begin{abstract}
Extreme-mass-ratio inspirals (EMRIs) are one of the primary targets for the recently adopted millihertz gravitational-wave (GW) observatory LISA. Some previous studies have argued that a fraction of all EMRIs form in matter-rich environments, and can potentially explain the dozens of soft X-ray band ($\sim 10^{-1} \rm keV$), low-frequency ($\sim 0.1$ mHz) periodic phenomena known as quasi-periodic eruptions (QPEs) and quasi-periodic oscillations (QPOs). Here, using a representative EMRI population retrofitted with cutoffs on LISA-band SNRs and luminosity distances to account for the sensitivity of current instruments, we estimate the mean frequency band in which QPEs and QPOs originating from detectable LISA EMRIs may be emitting an X-ray signal ``today'' (i.e., in 2024) to be $0.46 \pm 0.22$ mHz. We also model the well-known QPO source, RE J1034+396, which falls in this frequency band, as an EMRI assuming its primary black hole mass to be $10^6-10^7 M_\odot$. Through a prior-predictive analysis, we estimate the orbiting compact object's mass to be $46^{+ 10}_{-40} M_\odot$ and the source's LISA-band SNR as $\approx 14$, highlighting it as a candidate multi-messenger EMRI target. We also highlight the role of current and near-future X-ray and UV observatories in enabling multi-messenger observations of EMRIs in conjunction with LISA, and conclude with a discussion of caveats of the current analysis, such as the exclusion of eccentricity and inclination from the model, and the measurability of sub-solar mass compact object EMRIs.
\end{abstract}

\begin{keywords}
black hole physics
--
accretion, accretion discs
--
gravitational waves
--
transients: black hole mergers
--
transients: black hole - neutron star mergers
--
galaxies: nuclei
\end{keywords}



\section{Introduction}\label{sec:introduction}
Extreme-mass-ratio inspirals (EMRIs) are black hole binary systems in which a stellar-mass compact object (CO) of mass $\mu \sim 1-10^2 M_\odot$ completes $\sim 10^5$ orbits around a supermassive black hole (MBH) of mass $M \sim 10^5-10^7 M_\odot$ in its strong-gravity regime $(r \sim 10~GM/c^2)$ over a period of $\sim 1-10$ years. They emit millihertz gravitational waves signal (GW) in the process \citep{Amaro_Seoane_2015, Barack_2004}, and are one of the main targets of the recently adopted European Space Agency (ESA)-led space-based GW observatory, Laser Interferometer Space Antenna (LISA)~\citep{consortium2013gravitational,amaroseoane2017laser,2024arXiv240207571C}. EMRIs have the potential to be  unique scientific laboratories for conducting tests of general relativity (GR), constraining the astrophysical MBH mass-function, and inferring cosmological parameters \citep{Amaro_Seoane_2007,Amaro_Seoane_2018,Gair_2013,Babak_2017,berry2019unique,Laghi_2021}.\\

Over the years, various EMRI formation channels have been proposed~\citep{Amaro_Seoane_2018}: (i) Direct capture, where the many-body scattering in the nuclear star cluster around an MBH pushes a CO into the gravitational-wave loss cone \citep{HilsBender1995,SigurdssonRees1997};
(ii) The tidal disruption of a CO binary by the MBH \citep[Hills mechanism; ][]{1988Natur.331..687H} where one component is ejected as a hypervelocity star and the other component becomes more tightly bound to the MBH \citep{2005ApJ...631L.117M}; (iii) The capture of COs and their stellar precursors by the accretion disk in an active galactic nucleus (AGN) and their subsequent migration to tighter orbits around the MBH due to disk-CO interactions \citep{1978AcA....28...91P,SyerClarkeRess1991,Rauch1995,KarasSubr2001,Pan_2021}; and (iv) The formation of stellar objects later evolving into COs {\em in situ} within the accretion disk and their continual migration towards the MBH while embedded within the disk \citep{1978AcA....28...91P,1999Ap&SS.265..501C,2003ApJ...590L..33L,Levin2007,SiglSchnittmanBuonanno2007}. While the disk-assisted formation channels (iii)-(iv) are only relevant for the $\sim 10^{-2}-10^{-1}$ fraction of galactic nuclei that are (or recently have been) AGN, in some population models such as in that of \citet{Pan_2021} they have significantly higher formation rates than the ``dry'' channels (i)-(ii) and, consequently, produce a major fraction of the predicted LISA EMRIs. 

In addition to GW radiation, EMRI sources formed in channels (iii)-(iv) can trigger high energy electromagnetic (EM)-band emissions owing to CO interactions with the matter in the MBH accretion disk, making them viable candidates for multi-messenger detections \citep{LehtoValtonen1996,Semerak1999,Ivanov1998,2016MNRAS.457.1145P, Sukov__2021}. Channels (i)-(ii) can also lead to CO -- hot flow interactions in low-luminosity galactic nuclei. However, because of lower gas densities, the emerging weak EM signal limits such GW sources to the local Universe and rather fine-tuned setups with a low likelihood, such as a pulsar on a tight orbit around Sgr~A* \citep{2017FoPh...47..553E}. 

Identification of individual multi-messenger sources would be invaluable for simplifying data analysis schemes such as the LISA global-fit \citep{Vallisneri_2009,PhysRevD.72.043005,PhysRevD.107.063004}.  This can be done by setting tight constraints on the extrinsic parameter set, while lifting degeneracies arising from the GW-only inference of beyond vacuum-GR EMRI parameters~\citep{Speri:2022upm, kejriwal2023impact}. Such observations will also enable independent constraints on the redshift of the host galaxy and the properties of the galactic environment, without invoking statistical methods like the dark siren method (\citeauthor{1986Natur.323..310S}, \citeyear{1986Natur.323..310S};  \citeauthor{PhysRevD.86.043011}, \citeyear{PhysRevD.86.043011}; see also \citeauthor{gray2023joint}, \citeyear{gray2023joint} for a review). Multi-messenger detections of EMRIs would thus help to decouple the MBH parameters from different galaxy components \citep{KormendyHo2013, 2021bhns.confE...1K} and significantly improve the constraining power of LISA on the luminosity distance-redshift relation in cosmology \citep{Schutz1986,HolzHughes2005}. The precise knowledge of the parameters of an accreting MBH would additionally allow the resolution of long-standing conundrums such as the radio loud/quiet dichotomy of AGN \citep{Sikora2007} and their conjectured relation to MBH spin \citep[see, e.g.,][]{Tchekhovskoy2010}.

Independent of the specific model, one can expect the CO-disk interactions to manifest as regular time variability in the optical to X-ray bands, corresponding to the temperatures of the accretion disk at various distances from the MBH \citep{ShakuraSunyaev1973}. Another possible signature could come in the X-ray/radio band in the case of launched outflows \citep{Sukov__2021,Li_2023,2024arXiv240210140P}. Variability in the soft X-ray band (around $\sim 10^{-1}\ \rm keV$ or $\sim 10^6 \,\rm K$) corresponds to regions in the disk within a few gravitational radii near the MBH. At such distances, GW radiation is also non-negligible. The characteristic period of this variability should then correspond to the orbital period of the CO inferred from the GW signal of the source, which will be $\sim$ hours to minutes for EMRIs relevant to LISA (see Section \ref{sec:backevolve}).

In line with this description, the recently discovered quasi-periodic eruptions (QPEs), which are signals of semi-regular bursts in soft X-ray energy bands around $\sim 10^{-1} \rm keV$, and which can be traced back to centers of known galaxies, have been proposed to be EM counterparts of environment-rich EMRIs \citep{Sukov__2021, Chen_2022, Metzger_2022, 2024arXiv240210140P, linial2023emri, Franchini_2023, zhou2024probing}. Notably, QPEs may also be explained by accretion disk instabilities around the MBH \citep{Miniutti2019} or gravitational lensing during the inspiral of massive black hole binaries (MBHBs) \citep{2021MNRAS.503.1703I}, but such models only explain a subset of the observed signals. Coincidentally, some QPEs have been found in galaxies hosting MBHs of masses $\sim 10^6 M_{\odot}$ \citep{Wevers_2022, King_2023}, which is consistent with the mass range of expected LISA-relevant EMRIs ($10^5 - 10^7 M_\odot$) further motivating a multi-messenger search.

Another type of possibly related phenomenon are the less abrupt quasi-periodic oscillations (QPOs) in the X-ray fluxes of AGNs, such as the ones reported in RE J1034+396 \citep{Gierliński2008}, 2XMM J123103.2+110648 \citep{Lin_2013}, Sw J1644+57 \citep{Wang_2014}, MS 2254.9-3712 \citep{Alston_2015}, 1H 0707-495 \citep{2016ApJ...819L..19P}, Mrk 766 \citep{Boller2001,2017ApJ...849....9Z}, ASASSN-14li \citep{doi:10.1126/science.aar7480} or MCG-06-30-15 \citep{Gupta2018}, all with the estimated periods of oscillations falling within few hours. While many studies have tried to make a connection between AGN QPOs and their well-known kHz analogues in X-ray binaries \citep{2000ARA&A..38..717V,2006ARA&A..44...49R,Ingram_2019}, their mechanisms may be completely unrelated. Indeed, it has been suggested in the literature that AGN QPEs and QPOs may originate from the same source under different configurations \citep{10.1093/mnrasl/slad052, linial2023emri}. 

Hence, in this paper, we proceed with the assumption that AGN QPEs and QPOs are both caused by EMRIs, jointly referring to them as quasi-periodic phenomena (QPPs). The relative differences in the intensity of peaks and duty cycles between QPEs and QPOs will then be assumed to be caused by variations in the particular parameters of the sources such as the orbital inclination of the EMRIs or the accretion states of the disks (see Section \ref{sec:eccinc} for a detailed discussion).

\subsection{Executive Summary of Results}

An effective multi-messenger study of LISA-relevant EMRIs would require us to first determine the QPP frequency band in which their EM counterparts will most likely be observable. We find in Section~\ref{sec:backevolve} that EMRIs in the LISA observation window (assuming an operation time from the year 2037 to 2041) should emit QPP signals ``today'', i.e. in 2024, in a wide frequency range with a mean of $\approx 0.46$ mHz and a standard-deviation of $\sigma \approx 0.22$ mHz. As a demonstrative case study, in Section~\ref{sec:REJ1034396} we analyze the well-known QPO source RE J1034+396, which has a suitable oscillatory signature with the period of $\sim 1 $ hr in the relevant EM energy band of $\sim 0.2-10 \, \rm keV$ \citep{Gierliński2008, Jin_2020}. Assuming an MBH mass of $10^6-10^7 M_\odot$, we model it as an EMRI to obtain prior-predictive distributions on the CO's mass of $46^{+10}_{-40}M_\odot$, the source's LISA-band SNR with a mean of $\approx 14$, and LISA-band detection probability of $\approx 0.25$ with an SNR cutoff of $15$, highlighting it as a potential multi-messenger EMRI candidate. Given the QPP frequency band, in Section~\ref{sec:targetedsearch} we then highlight some current and near-future X-ray observatories that are sufficiently sensitive to signals in this band, and can thus be employed to obtain a catalog of EM counterparts of LISA-relevant EMRIs. We discuss some caveats of our analysis in Section~\ref{sec:caveats}, and conclude in Section~\ref{sec:conclusion}. We assume the standard flat $\Lambda$CDM model with $H_0=70\,{\rm km\,s^{-1}\,Mpc^{-1}}$ and $\Omega_{\rm m}=0.3$ throughout the text.
\section{QPP Frequencies of LISA-relevant EMRIs Today}\label{sec:backevolve}
\subsection{EMRI catalog}\label{sec:EMRIcatalog}
We begin by assuming that the astrophysical EMRI population is described by the \texttt{M1} population catalog of \cite{Babak_2017}\footnote{A more complete analysis may benefit from analyzing all 12 EMRI catalogs (\texttt{M1-M12}) generated in \cite{Babak_2017} to account for uncertainties pertaining to their underlying astrophysical models. Our study illustratively examines only the fiducial \texttt{M1} model.}. However, the source parameters in the \texttt{M1} catalog are given at plunge, i.e. when the CO is near the MBH's innermost stable orbit. \cite{pozzoli2023computation} back-evolved these plunge-time parameters to a random time $t \in [0, T_{\rm back}]$ years before plunge (where $T_{\rm back} = 20(M/10^4M_\odot)$ years) using the leading-order Peters \& Mathews waveform model \citep{1963PhRv..131..435P}, given by:
\begin{align}
    &\frac{\mathrm{d}f_{\rm orb}}{\mathrm{d}t} = \frac{96G^{5/3}}{5c^5}\left(2\uppi\right)^{8/3}\mathcal{M}^{5/3}f^{11/3}_{\rm orb}\mathcal{F}(e),\label{eq:BonettiSesanna1}\\
    &\frac{\mathrm{d}e}{\mathrm{d}t} = -\frac{G^{5/3}}{15c^5}\left(2\uppi\right)^{8/3}\mathcal{M}^{5/3}f^{8/3}_{\rm orb}\mathcal{G}(e),\label{eq:BonettiSesanna2}
\end{align}
where $f_{\rm orb}$ is the orbital frequency of the EMRI, $e$ is its eccentricity, and
\begin{align}
    &\mathcal{M} = \frac{\left(M\mu\right)^{3/5}}{\left(M+\mu\right)^{1/5}},\label{eq:BonettiSesanna3}\\
    &\mathcal{F}(e) = \frac{1+(73/24)e^2+(37/96)e^4}{\left(1-e^2\right)^{7/2}},\label{eq:BonettiSesanna4}\\
    &\mathcal{G}(e) = \frac{304e+121e^3}{\left(1-e^2\right)^{5/2}}.\label{eq:BonettiSesanna5}
\end{align}
\cite{pozzoli2023computation} thus describes a more realistic EMRI catalog such that only a fraction of EMRIs will plunge in the LISA observation window. We employ this modified catalog in our study. Additionally, note that the purpose of the \texttt{M1} catalog is to synthesize a generous set of EMRIs that is significantly larger than what will be detected by LISA, which is then used to explore its detection capabilities. As such, $T_{\rm back}$ for $M \sim 10^6 M_\odot$ can be as large as $\sim 1000$ years. Sources this far away from plunge will produce negligible $(\ll 1)$ SNRs in the LISA band in general. For simplicity, we thus only consider sources that plunge within $t \leq 100$ years after the start of LISA observation window. We additionally require the sources to have eccentricities, $e \leq 0.8$, mass-ratios, $\mu/M \leq 10^{-4}$, and inclination angles between the orbital plane and accretion disk, $\iota \geq 0.01$. We assume the plane of the accretion disk is perpendicular to the spin-axis of the MBH, but briefly discuss the effect of misaligned planes in Section~\ref{sec:eccinc}. These preliminary cutoffs are based on conservative estimates on the distribution of environment-rich EMRI parameters and validity of the GW waveform model (see also discussion in Sections~\ref{sec:REJ1034396} and \ref{sssec:co_mass}).

Equations ~\eqref{eq:BonettiSesanna1}-\eqref{eq:BonettiSesanna5} do not account for contributions from the CO-disk interactions, due to the unavailability of such models for generic orbit (eccentric and inclined) Kerr MBH EMRIs. While their inclusion may introduce significant dephasing in the long-term inspiral evolution, they are only expected to contribute to the inspiral at the next-to-leading order, as shown by previous studies \citep[see e.g.,][and also subsection \ref{sssec:co_mass}]{2011PhRvD..84b4032K,2014PhRvD..89j4059B, Speri:2022upm}. However, we stress that these effects may significantly influence the long-term evolution of the EMRI and should be explored in future studies. A more realistic analysis should also use higher-order inspiral evolution models with a larger parameter space (including, for example, the evolution of inclination, MBH spin, etc.), in contrast to the leading post-Newtonian (PN) order Peters \& Mathews Schwarzschild EMRI model employed here for its computational feasibility. 
\subsection{Back-evolving to current orbital frequencies}
\begin{figure}
\centering
\includegraphics[width=0.45\textwidth]{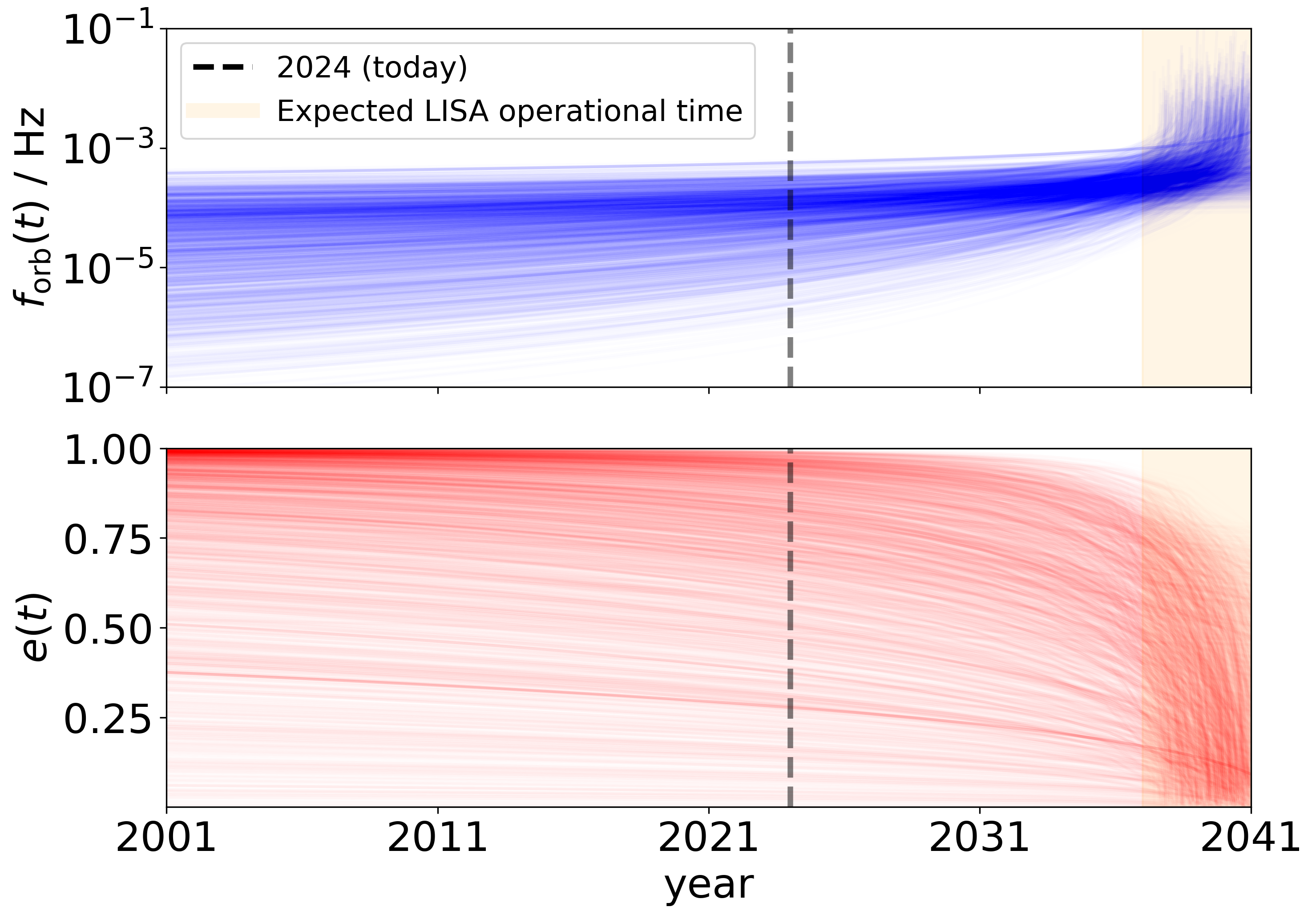}
\caption{Back evolution of the trajectories of detector-frame orbital frequency, $f_{\rm orb}$, (upper panel) and eccentricity, $e$, (lower panel) for EMRIs that are uniformly distributed in LISA's observation window (orange band), i.e. years 2037-2041 to 2001. Time ``today'', 2024, is demarcated by the black dashed line. The boldness of each trajectory is proportional to the EMRI SNR in the LISA observation time calculated using the approximate model described in the text.}
\label{fig:evolution}
\end{figure}

We randomly distribute the \texttt{M1} catalog parameters in the expected LISA operation window (Gregorian calendar year 2037 to 2041)\footnote{\hyperlink{https://sci.esa.int/web/lisa/-/61367-mission-summary}{https://sci.esa.int/web/lisa/-/61367-mission-summary}}, and back-evolve the EMRI's orbital frequency and eccentricity parameters to the year 2024, i.e. ``today'', using Eqs. \eqref{eq:BonettiSesanna1} and \eqref{eq:BonettiSesanna2} (see Fig.\ref{fig:evolution}). From the EMRI orbital frequencies ``today'', $f_{\rm orb, init}$, we can then calculate the distribution of QPP frequencies for the population's EM counterparts, $f_{\rm QPP}$, as
\begin{align}
    f_{\text{QPP}} \equiv n_{\rm col} \times f_{\rm orb, init}\label{fqpoqpe}
\end{align}
where $n_{\rm col}$ is the number of CO-disk collisions per orbit, assumed to be 2 throughout our analysis and supported by previous estimates~(\citealt{linial2023emri, linial2023emri_UV}, see also \citealt{Sukov__2021}, \citealt{2021ApJ...921L..32X}, and \citealt{Franchini_2023}).  Note that this relation is only valid for EMRIs evolving in circular orbits, and while environment-rich EMRIs with COs evolving in the disk are expected to rapidly circularize under the effect of CO-disk interactions (\citeauthor{2004ApJ...602..388T}, \citeyear{2004ApJ...602..388T}; see also \citeauthor{Kley_2012}, \citeyear{Kley_2012}), such interactions will be small in EMRIs with misaligned accretion disk and CO orbital planes. While we use the approximate relation in Eq.~\eqref{fqpoqpe} in our study as a leading-order estimate, we emphasize that such a relation should be generalized to include eccentric EMRIs in future work. We also discuss the potential impact of introducing non-negligible eccentricity and inclination in the model in Section \ref{sec:eccinc}.

\begin{figure}
\centering
\begin{subfigure}{0.45\textwidth}
    \includegraphics[width=\textwidth]{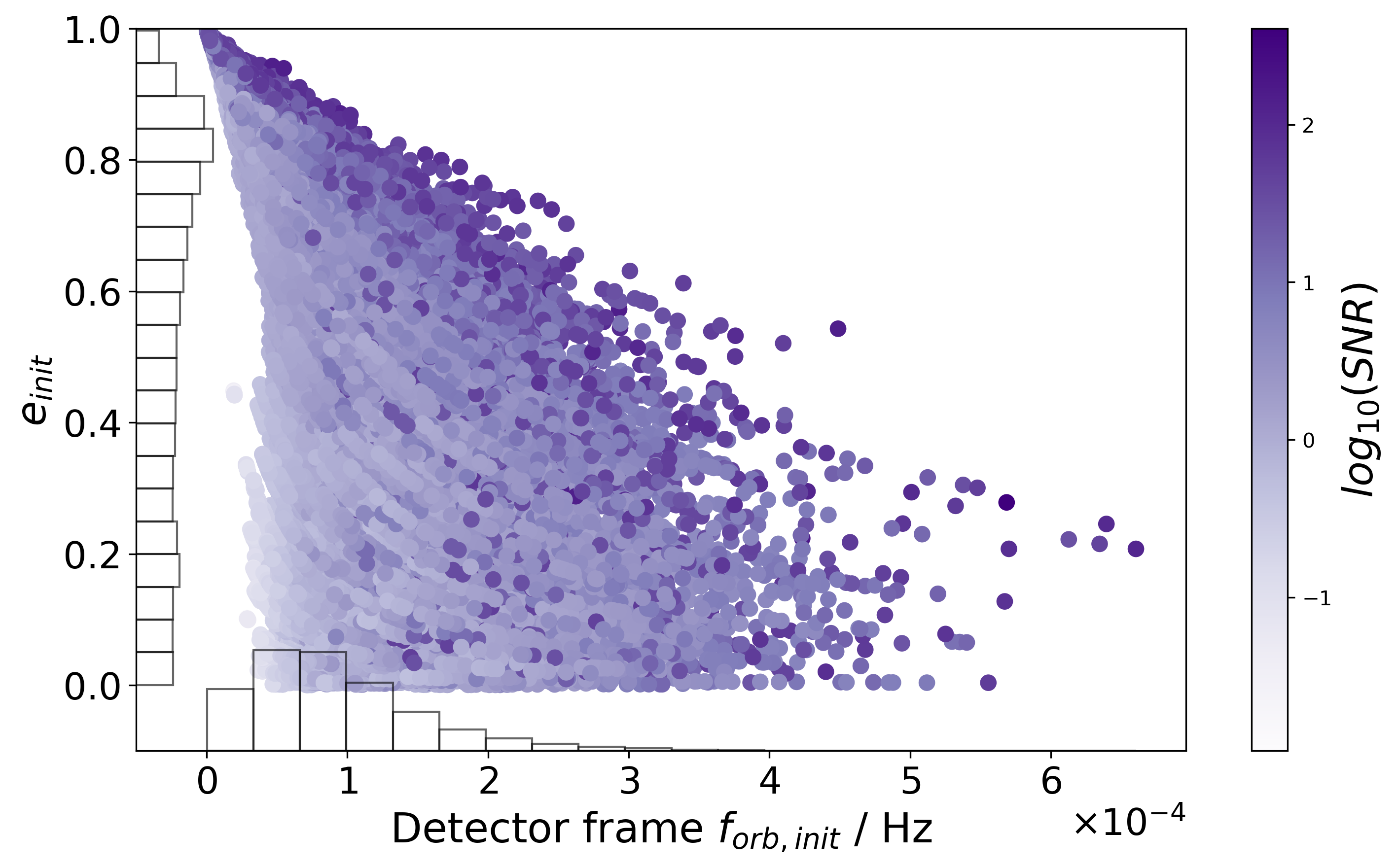}
\end{subfigure}
\begin{subfigure}{0.45\textwidth}
    \includegraphics[width=\textwidth]{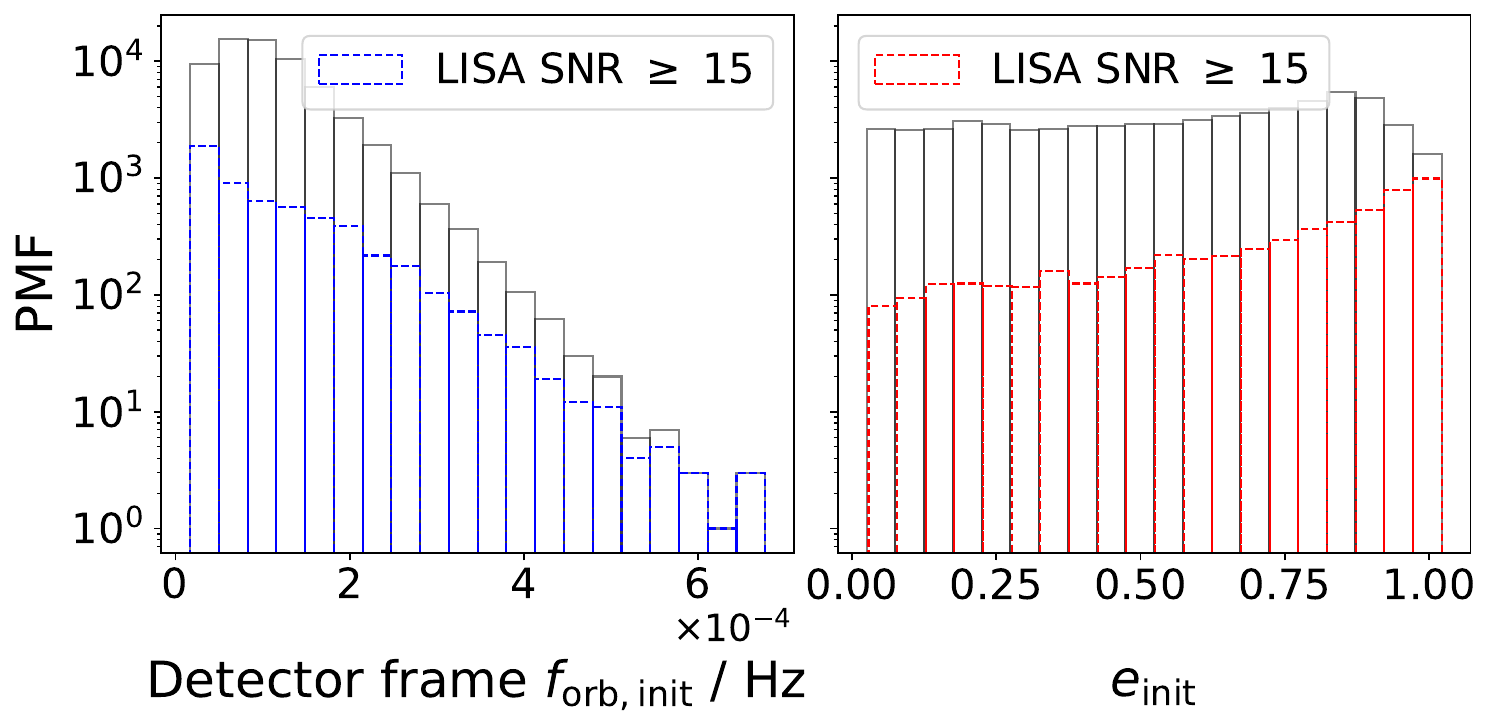}
\end{subfigure}
\caption{\justifying The joint distribution (\textit{top panel}) and 1-D marginals (\textit{bottom panel}) of the detector-frame orbital frequency, $f_{\rm orb,init}$ and eccentricity $e_{\rm init}$ of LISA-band EMRIs ``today'', i.e. in 2024. Each marker in the top panel is darkened in proportion to the source's SNR. In the bottom panel, the distributions of sources satisfying the LISA-band detectability cutoff based on these SNRs are also depicted.}
\label{fig:initdistribution}
\end{figure}

In Fig. \ref{fig:evolution}, we visualize the back-evolved detector-frame orbital frequency and eccentricity trajectories for the considered EMRI population in the LISA band till 2001.
We also estimate their LISA-band SNRs, $\rho$, using the solar system barycenter frame point-detector approximation of \cite{Bonetti_2020} (see Eq. (17), (18), and (23) in their paper).
The cross-section of these trajectories evaluated for the year 2024 thus gives the distribution of detector-frame orbital frequencies $f_{\rm orb,init} = f_{\rm orb, source}/(1+z)$ and eccentricities $e_{\rm init}$ ``today'', and are summarized in Fig. \ref{fig:initdistribution}. 

\subsection{Simultaneous EM and GW observability}

\begin{table*}
\begin{center}
    \begin{tabular}{|c|c|c|c|c|c|c|}
        \hline
        Name & Type & ref. & $z$ & $L_{x}$ & $f$  & MBH mass range \\
            &   &   &   &    (erg/s)    & ($\times 10^{-4}$ Hz) &   ($\times 10^6 M_\odot$)\\
        \hline
        \hline
        GSN 069 & QPE & \cite{Miniutti2019} & 0.0181 & $1.1\times10^{43}$ & 0.3 & 0.4 - 2\\ 
        eRO-QPE1 & QPE & \cite{Arcodia2021} & 0.0505 & $3.3\times10^{42}$ & 0.2 & 0.1 - 1\\ 
        eRO-QPE2 & QPE & \cite{Arcodia2021} & 0.0175 & $1.0\times10^{42}$ & 1.2 & 0.1 - 1\\ 
        eRO-QPE3 & QPE & \cite{2024arXiv240117275A} & 0.024 & $4.2\times 10^{41}$ & 0.1 & 0.9 - 5.3\\
        eRO-QPE4 & QPE & \cite{2024arXiv240117275A} & 0.044 & $1.27 \times 10^{43}$ & 0.2 & 17 - 68\\ 
        RX J1301.9+2747 & QPE & \cite{Giustini_2020} & 0.0237 & $1.6\times10^{42}$ & 0.6 & 0.8 - 2.8\\ 
        RE J1034+396 & QPO & \cite{Gierliński2008} & 0.042 & $4.0\times10^{43}$ & 2.8 & 1 - 10\\
        2XMMJ123103,2+110648 & QPO & \cite{Lin_2013} & 0.13 & $2.5\times10^{42}$ & 0.7 & 0.09 - 1.1\\
        ASASSN-14li & QPO & \cite{doi:10.1126/science.aar7480} & 0.0206 & $2.0\times10^{43}$ & 76.3 & 0.6 - 12.5\\
        ASASSN-20qc & QPOut$^{*}$ & 
        \cite{2024arXiv240210140P} 
        & 0.0136 & $(2-4)\times 10^{44}$ & 0.019 & 16-160 \\
        Swift J0230+28 & QPE + QPOut(?)$^{*}$ & 
        \cite{2023arXiv230903011G} 
        & 0.0053 & $(3-6)\times 10^{42}$ & 0.019 & 1.6-10 \\
        \hline
    \end{tabular}
    \caption{Summarized properties of QPEs and QPOs discovered so far in the frequency band of $\sim 0.1$ mHz. From left to right, we specify the source's name, type (QPO/QPE/QPOut), redshift ($z$), peak soft X-ray (0.2-2 keV or 0.2-10 keV) luminosity, the QPP frequency ($f$), and current lower and upper bounds on the source's MBH mass. 
    Asterisk ($^{*}$): QPOuts stands for the newly-discovered phenomenon of quasi-periodic outflows where the X-ray modulation is driven by a narrow bandpass dominated by an outflow. Recurrent spectral features reminiscent of outflows were reported from Swift J0230+28, but their statistical robustness is yet to be verified. }
    \label{tab:QPsummary}
\end{center}
\end{table*}

The \texttt{M1} catalog of \cite{Babak_2017} has EMRIs at redshifts up to $z = 4.5$ or luminosity distance $d_L \sim 40$ Gpc, which is much larger than the distances covered by current state-of-the-art X-ray band instruments. The observable volume is further restricted by the requirement of small integration times necessary for capturing the $\sim$ hours variability of relevant QPP sources. Similarly, for LISA-band EMRI observations, detectability of sources will be constrained by their GW SNRs. A large fraction of EMRIs in the \texttt{M1} catalog have SNRs $\leq 1$ even after setting the cutoffs mentioned in Section~\ref{sec:EMRIcatalog}, making them undetectable. For a simultaneous observation in EM and GW bands, we thus must set realistic cutoffs on the observable cosmological volume (or equivalently, the maximum luminosity distance, $d_L$) and the lowest detectable LISA-band SNR, $\rho$.

The luminosity distance cutoff can be set by considering both the sensitivity threshold of current X-ray instruments and the required exposure time to significantly detect the QPP EM emission. For the mean X-ray luminosity of QPE sources, $\overline{L}_{\rm X}=2\times10^{42}\,{\rm erg\,s^{-1}}$ (see Table~\ref{tab:QPsummary}), a luminosity distance of $d_{\rm L}\lesssim 1\,{\rm Gpc}$ results in the X-ray flux density threshold of $f_{\rm X, sens}\equiv L_{\rm X}/4\pi d_{\rm L}^2\gtrsim 1.7\times 10^{-14}\,{\rm erg\,s^{-1}\,{\rm cm^{-2}}}$. Since current X-ray detectors (Swift, XMM-Newton, and eROSITA), meet the sensitivity threshold $f_{\rm X, sens}$ for suitable integration times of a few tens of minutes, we treat the luminosity distance $d_L = 1$ Gpc as the cutoff (see also Section~\ref{sec:targetedsearch} for a more detailed discussion of the capabilities of current and future X-ray detectors).

For the LISA-band SNR cutoff, $\rho$, we choose a value of 15, given the unavailability of a universally accepted threshold (which should scale with the information volume of the signal space \citep{Moore_2019}). Our specific choice is motivated by the estimates produced as part of the mock LISA data challenges (MLDC), which hinted at an EMRI measurability threshold of $\rho \approx 20$ for a two-year observation window~\citep{Babak_2010}, and accounting for the accumulation of SNR over the full LISA observation window, assumed here to be four years \citep{Amaro_Seoane_2021}. The SNR cutoff for an EMRI's detectability would also depend on the number of parameters under simultaneous inference. If, for example, an EM counterpart detection tightly constrains some extrinsic EMRI parameters, the SNR cutoff can be expected to further drop, justifying our choice of $\rho_{\rm cutoff} = 15$. However, this cutoff ignores the effect of data gaps and glitches in the LISA data stream which complicates inference and can thus potentially increase the SNR requisites. In general, their contributions are expected to be negligible in the inference of LISA-relevant sources \citep{Dey2021, PhysRevD.100.022003} and are thus ignored in our analysis.

\begin{figure}
\centering
\includegraphics[width=0.455\textwidth]{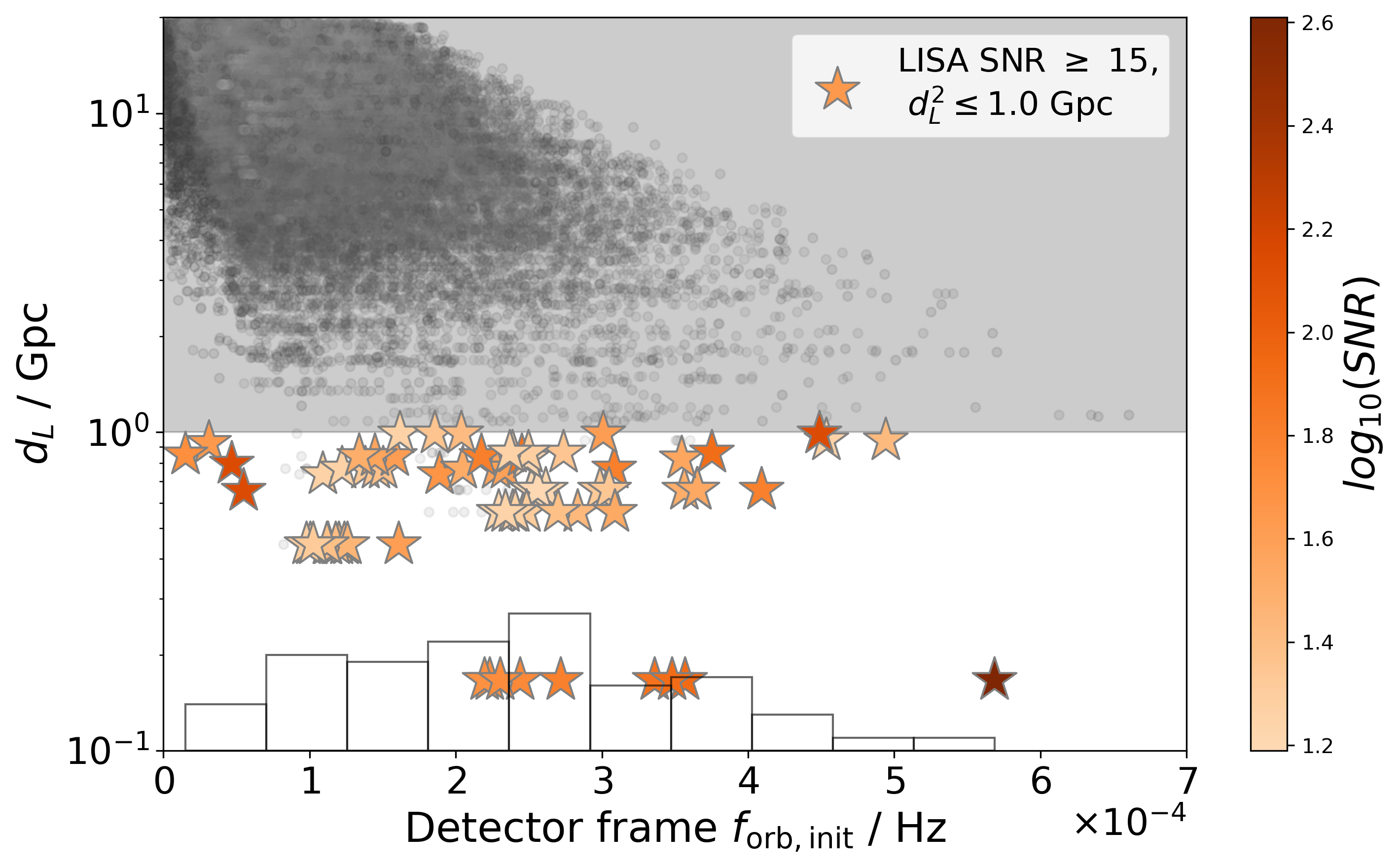}
\caption{Distribution of EMRI sources with respect to detector-frame $f_{\rm orb,init}$ and their luminosity distance from the detector, $d_L$, ``today'', i.e. in 2024. Sources in the grey band have $d_L \geq 1$ Gpc. The stars mark all sources that simultaneously satisfy $d_L \leq 1$ Gpc and the cutoff on the LISA-band SNR, $\rho \geq 15$. 1D marginal distribution of the detector frame  $f_{\rm orb, init}$ of sources that satisfy the cutoffs is depicted along the horizontal axis, which thus describes the distribution of orbital frequencies of candidate multi-messenger EMRIs.}
\label{fig:JointObservation}
\end{figure}

Fig. \ref{fig:JointObservation} describes the distribution of the detector-frame orbital frequencies and eccentricities ``today'' (in 2024) after passing the EMRI population through a cutoff of $d_{\rm L} \leq 1$ Gpc and $\rho \geq 15$.\footnote{Note that the large gap along the $d_L$-axis in Fig.~\ref{fig:JointObservation} between the sources marked with a star is an artifact of the sampling resolution of Babak et al. \citep{Babak_2017}'s population, but since we are interested in the span of these sources along the frequency axis, this artifact can be safely ignored.} The corresponding 1-D sample distribution of $f_{\rm orb, init}$ has mean $\langle f_{\rm orb, init}\rangle\approx 2.3  \times 10^{-4}$ Hz and 1$\sigma$ standard-deviation of $1.1 \times 10^{-4}$ Hz. Using Eq. \ref{fqpoqpe} with $n_{\rm col} = 2$, a rough estimate on the frequency of LISA-relevant EMRI QPPs is obtained as $\langle f_{\rm QPP} \rangle \approx 4.6 \pm 2.2 \times 10^{-4}$ Hz. This estimate represents a generous spread in the $\sim 0.1 $ mHz frequency band and numerous candidate and confirmed QPEs and QPOs in this frequency band have been reported in the literature \citep{Miniutti2019,Arcodia2021,2024arXiv240117275A, Giustini_2020,Chakraborty_2021,Gierliński2008,Jin_2020,Alston_2015,Lin_2013,Boller2001,2016ApJ...819L..19P,doi:10.1126/science.aar7480,Wang_2014, Gupta2018}. However, follow-up studies have speculated upon the statistical robustness of some of these observations (see for example \cite{2005A&A...431..391V} and~\cite{,2023ApJ...946...52Z}). 

A list of QPPs that fall near the proposed frequency band, and with statistically significant observations, is presented in Table~\ref{tab:QPsummary}. For completeness, we additionally list a new class of nuclear transients called QPOuts (quasi-periodic outflows) discovered by \citet{Pasham_2024} which are also proposed to be (early) electromagnetic counterparts of large mass-ratio inspirals. In \citet{Pasham_2024}, the authors observed quasiperiodic dips in the X-ray band caused by the enhanced absorption due to an intermittent fast outflow. The QPOut was interpreted as possibly caused by an intermediate-mass black hole orbiting a supermassive black hole as supported by the simulations of \citet{Sukov__2021}. However, only one such confirmed transient has been observed so far (but see also \citealt{2023arXiv230903011G}) and is most likely far from merger. We thus leave its analysis for future studies.

As we discuss in Section \ref{sec:targetedsearch}, some current and near-future EM observatories should be able to detect QPP signals in the estimated frequency band, enabling potential multimessenger observations of LISA-EMRIs.

\section{Modeling RE J1034+396 as an EMRI}\label{sec:REJ1034396}

RE J1034+396, an AGN located at $z = 0.042$, is a confirmed QPO with a frequency of several times $10^{-4}$ Hz \citep{Alston_2014, Jin_2020}. Several estimates on the mass of its MBH are available, (\citeauthor{https://doi.org/10.1111/j.1365-2966.2009.15662.x}, \citeyear{https://doi.org/10.1111/j.1365-2966.2009.15662.x}; \citeauthor{10.1093/mnras/sty1366}, \citeyear{10.1093/mnras/sty1366}; see also \citeauthor{2016A&A...594A.102C}, \citeyear{2016A&A...594A.102C} for the comparison of the inferred MBH mass using different methods), but it is generally agreed to be in the range of $10^6 - 10^7 M_\odot$. It has been observed multiple times over the span of about a decade (2007-2018) --- In 2007, the QPO's time period was observed to be $T_{2007} \sim 3730 \pm 80 s$ \citep{Gierliński2008} which in 2018 reduced to $T_{2018} \sim 3550 \pm 60 s$ \citep{Jin_2020}. If RE J1034+396 is modeled as an EMRI, the reduction in periodicity can be associated with GW-induced radiation reaction of the system, and/or loss of energy via CO-disk interactions. However, according to Table 1 in \cite{Jin_2020} \citep[see also Table 2 in][]{Alston_2014}, the QPO frequency fluctuated between $0.25 - 0.27$ mHz during 2007 - 2011 without any secularly evolving features. This can still be explained under the environment-rich EMRI model through eccentric orbit CO-disk interactions, in which the torque forces generated by the disk's density wave perturbations both inside and outside of the CO orbit compete with each other, sometimes canceling each other out to create ``migration traps'', or even exerting a net-outwards force on the CO \citep{2016ApJ...819L..17B, Secunda_2020}. A detailed analysis of this effect, however, is beyond the scope of our study, and we assume a secular drift between the reported QPO time-periods in 2007 ($3730 \pm 80$s) and 2018 ($3550 \pm 60$s) in our model, which mark the earliest and latest reported values of the QPO in RE J1034+396.

\subsection{CO mass estimate for RE J1034+396}

Assuming two collisions per orbit, from Eq.~\eqref{fqpoqpe}, the corresponding orbital frequencies of the EMRI source in 2007 and 2018 are
\begin{align}\label{frequencies}
    f_{{\rm orb},i} = \frac{1}{2T_i} = \begin{cases}
        1.34 \times 10^{-4} \, \rm{Hz} & i = 2007\\
        1.41 \times 10^{-4} \, \rm{Hz} & i = 2018
    \end{cases}
\end{align}
such that the observed rate of change of the orbital frequency, $\dot{f}_{\rm orb, obs}$, can be approximated to linear-order as
\begin{align}
\dot{f}_{\rm orb,obs} \approx \frac{f_{\rm orb,2018}-f_{\rm orb,2007}}{2018-2007} \approx 6.18 \times 10^{-7} \ \ \text{Hz yr}^{-1}.
\end{align}

We can now use the distribution of $\dot{f}_{\rm orb,obs}$ to optimize the corresponding theoretical prediction, 
\begin{align}
    \dot{f}_{\rm orb,th} \equiv 
\dot{f}_{\rm orb,th}(M,\mu,a,e_0,\iota,f_{\rm orb,2007})
\end{align}
by minimizing the following $\chi^2$-error function,
\begin{align}\label{chi2error}
    \chi^2 \equiv \left(\dot{f}_{\rm orb,obs} - \dot{f}_{\rm orb,th}\right)^2.
\end{align}
Here $\mu$ is the unknown mass of the CO, $a$ is the dimensionless spin of the MBH, $e_0$ is the initial eccentricity of the CO's orbit, and $\iota$ is the initial inclination of the CO's orbital plane (which does not usually evolve significantly in the parameter regime we are considering). The inclusion of $\{a,e_0,\iota\}$ in the parameter set allows us to gauge their correlations with the calculated CO mass. A discussion on the choice of prior on $\{a,e_0,\iota\}$ and their effect on our analysis is presented in Appendix \ref{app:REJ}. We adopt $\dot{f}_{\rm orb,th}$ from the waveform model of \cite{Fujita_2020}, as implemented in the \texttt{5PNAAK} waveform model of the \texttt{FastEMRIWaveform (FEW)} toolkit~\citep{Chua:2020stf,PhysRevD.104.064047}.  Due to the unavailability of CO-disk interaction models for inclined and eccentric orbit EMRIs, we exclude their effects from our analysis. While these effects are subdominant in comparison to contributions from eccentricity and inclination, we stress that generic orbit environment-rich EMRI models will be necessary before drawing strong conclusions.

\begin{figure*}
\centering
\includegraphics[width=0.7\textwidth]{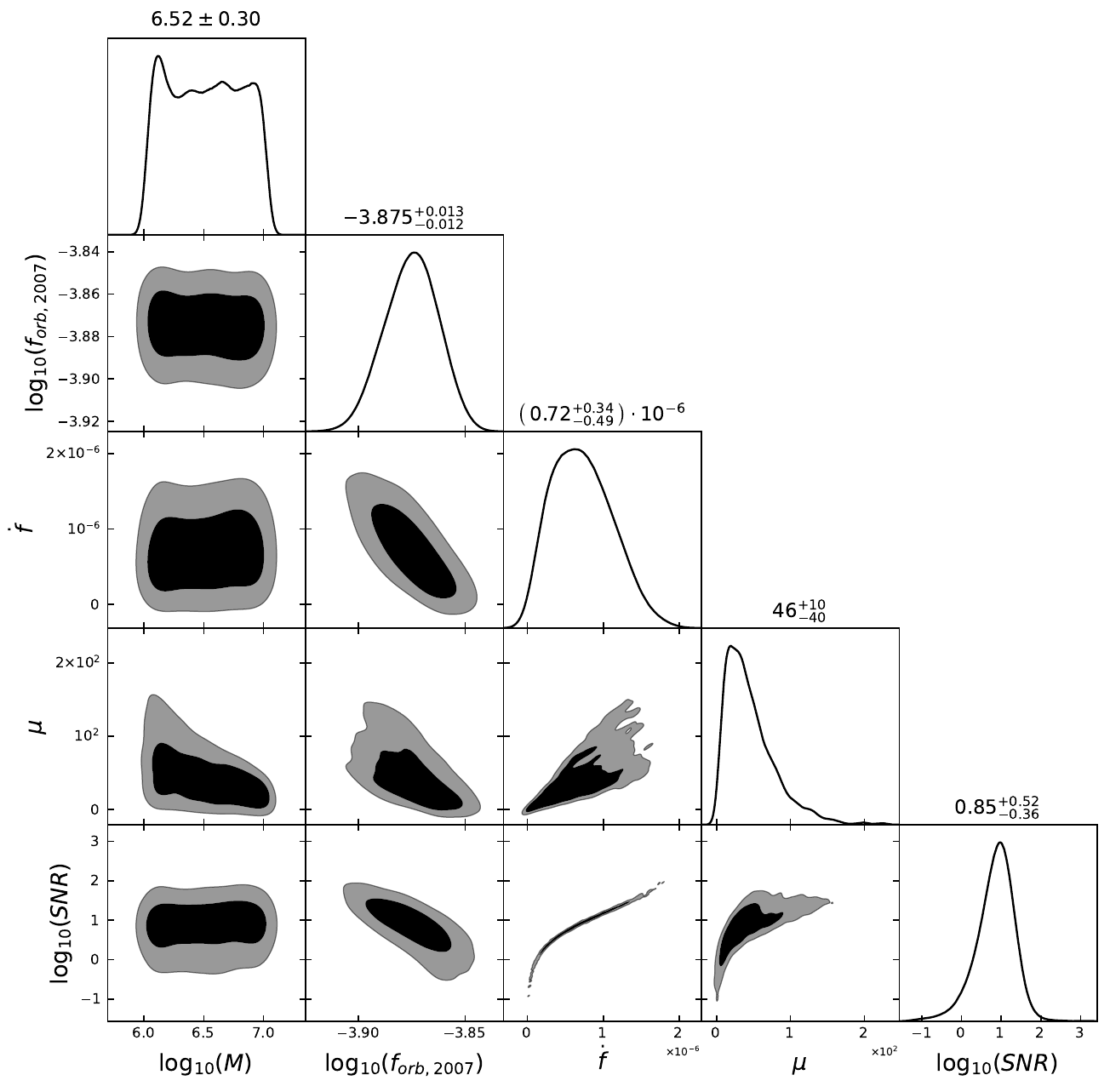}
\caption{Marginalized distributions over $\{M,f_{\rm orb,2007},\dot{f}_{\rm orb,obs},\mu,\text{SNR}\}$ of RE J1034+396 modeled as an extreme-mass-ratio inspiral. $\{M,f_{\rm orb,2007},\dot{f}_{\rm orb,obs}\} $ are estimated from current observational constraints on RE J1034+396. The thus obtained prior-predictive distribution on $\mu$ is described in the fourth panel, and the corresponding LISA-band SNR in the final panel. The full 8-dimensional distribution (including $\{a,e_0,\iota\}$) is available in Appendix \ref{app:REJ} as Fig. \ref{fig:SNRdist}.}
\label{fig:SNRdistmarginal}
\end{figure*}

The first three panels in Fig. \ref{fig:SNRdistmarginal} describe the distribution on the parameter set $\{M,f_{\rm{orb,2007}},\dot{f}_{\rm{orb,obs}}\}$, from which we infer the best-fit CO mass of $$\mu \approx 46_{-40}^{+10} M_\odot,$$ as depicted in the figure's fourth panel. This estimate comfortably falls in the range of CO masses expected for LISA-relevant EMRIs ($\approx 1-100 M_\odot$). We also refer to Fig.~\ref{fig:SNRdist} in the Appendix which explicitly shows the distribution with respect to all other considered parameters, i.e. $\{a,e_0,\iota\}$. Note, however, that the correlation coefficient, $\mathcal{C}_{ij} \equiv \sigma_{ij}/\sqrt{\sigma^2_{i}\sigma^2_j}$, between $\mu$ and $\{a,e_0,\iota \}$ is relatively small with $\mathcal{C}_{\mu a} \sim 0.01, \mathcal{C}_{\mu e_0} \sim -0.005$, and $\mathcal{C}_{\mu \iota} \sim 0.02$ which implies that the prior choice of $\{a,e_0,\iota \}$ does not significantly affect the distribution of $\mu$ in this model. $\mu$'s correlation with $\dot{f}_{\rm{orb,obs}}$ is highest at $\mathcal{C}_{\mu\dot{f}_{\rm{orb,obs}}} \sim 0.9$, while with the initial orbital frequency is also high at $\mathcal{C}_{\mu f_{\rm{orb,2007}}} \sim -0.76$. Comparatively, $\mu$ is moderately correlated with $M$, $\mathcal{C}_{\mu M} \sim -0.5$.

\begin{figure}
\centering
\includegraphics[width=0.4\textwidth]{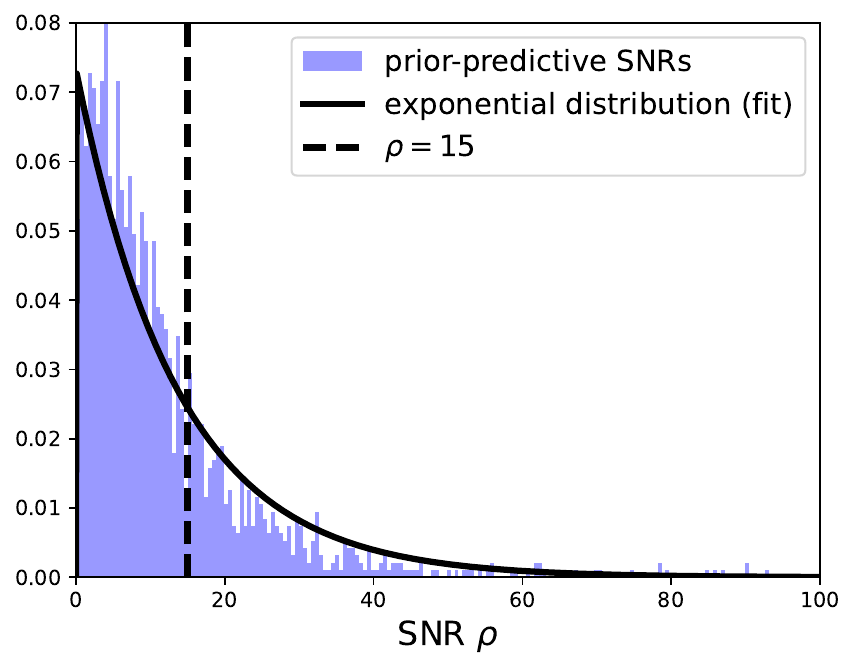}
\caption{Prior-predictive distribution of LISA-band SNRs of QPO RE J1034+396 modeled as an EMRI, with an indicative exponential fit. We assume the MBH mass to be in range $10^6-10^7 M_\odot$, mass of the orbiting body $\mu \approx 46^{+10}_{-40} M_\odot$, and conservative prior distributions on $\{a,e,\iota\}$ as described in the text. 
}
\label{fig:SNR_fit}
\end{figure}

\subsection{LISA band detectability of RE J1034+396}
The availability of the full EMRI parameter set, $(M,\mu,a,e_0,\iota,f_{\rm{orb,2007}})$, now allows us to evolve RE J1034+396 to LISA's operation time, 2037, and calculate the SNR, $\rho$, of the GW signal, given as \citep{Barack_2004}
\begin{align}
    \rho \equiv \sqrt{\left<h|h\right>}
\end{align}
where $h \equiv h(M,\mu,a,e_0,\iota,f_{\rm{orb,2007}},\boldsymbol{\psi}_{\rm{ext}})$ is the waveform template corresponding to the signal, and $\boldsymbol{\psi}_{\rm{ext}} = \{d_L = 0.18, \theta_S = 2.77, \phi_S = 0.69\}$ is the extrinsic parameter set defined by the luminosity distance $d_L$ and sky localisation parameters $\theta_S,\phi_S$ of the source. The one-sided detector noise weighted inner-product $\left<\cdot|\cdot\right>$ between two discrete time-domain signals $\alpha(t)$ and $\beta(t)$ is given as \citep{Chua_2022}:
\begin{align}
    \left<\alpha|\beta\right> \equiv 4\text{Re}\sum_\chi\sum_{f=0}^{f_N}\delta f\frac{\tilde{\alpha}^*_\chi(f)\tilde{\beta}_\chi(f) + \tilde{\alpha}_\chi(f)\tilde{\beta}^*_\chi(f)}{S_{{\rm n},\chi}(f)}.
\end{align}
Here $\tilde{\alpha}(f), \tilde{\beta}(f)$ are the frequency domain Fourier transformations of $\alpha(t), \beta(t)$, $f_N$ is the Nyquist sampling frequency, $\chi \equiv \text{I}, \text{II}$ are the two independent data-channels of LISA in its long-wavelength approximation (LWA) configuration, and $S_{\rm n}(f)$ is the power spectral density of LISA's noise channel $n_\chi$ evaluated at detector frequency $f$ \citep{Robson_2019}. 

The prior-predictive distribution of the source's LISA-band SNR is approximately exponential, with a sample mean of $\rho \approx 14$ (see Fig. \ref{fig:SNR_fit}). From this distribution, we can estimate the probability of detecting RE J1034+396 in the LISA band as ${\rm p}_{\rm detection}(\rho \geq 15) \approx 0.25$ for a LISA-band SNR cutoff of 15. Note that, in our model, the CO is on average separated from the MBH by about $18.8~GM/c^2$ when entering the LISA band and evolves to an average of $18.4~GM/c^2$ at the end of LISA observation period. If instead a model accounting for CO-disk interactions in generic orbits is used, the system will lose energy more rapidly and thus attain smaller separations when entering the LISA band, boosting the strength of its GW emission and hence its LISA-band SNR. Additionally, recall that the SNR cutoff of 15 was set based on conservative estimates for a GW-only inference. A multiband observation may enable the detection of sources with smaller SNRs. Illustratively, for smaller LISA-band SNR cutoffs, we have ${\rm p}_{\rm detection}(\rho\geq 12) = 0.33$ and ${\rm p}_{\rm detection}(\rho\geq 10) = 0.41$. These estimates suggest that RE J1034+396 may be a potential target for multimessenger EMRI observations.
\section{Discussion}\label{sec:discussion}

\subsection{Targeted QPP searches using current \& planned observatories} \label{sec:targetedsearch}
\subsubsection{Detectability threshold in soft X-rays}
For the electromagnetic detection of repeating X-ray transients, generally two aspects should be met: (i) sensitivity close to the eruption peak flux density and (ii) a sufficient EM-band SNR reached for the integration time that is shorter than the periodicity of a given source.

Conditions (i) and (ii) can be combined to estimate the minimal count rate required to detect the flare robustly to constrain its spectral energy distribution (SED), in particular its temperature. We consider the 5$\sigma$ threshold to be 25 counts. Since the QPE flares last for about an hour and repeat every few hours, the observation should last for at least a couple of hours, which results in the count rate of $25/7200\sim 0.003$ counts/s. For currently operational X-ray telescopes (except for Chandra which has a deteriorating soft X-ray sensitivity), this count rate translates into the limiting flux density threshold of $f_{X}\approx 6.3\times 10^{-14}\,{\rm erg\,cm^{-2}\,s^{-1}}$ for Swift XRT~\citep{2005SSRv..120..165B}, and $f_{X}\approx 3.2\times 10^{-15}\,{\rm erg\,cm^{-2}\,s^{-1}}$ for XMM/PN~\citep{2001A&A...365L..45H,2008A&A...479..283B} \footnote{Count to flux transformation is estimated using \url{https://heasarc.gsfc.nasa.gov/cgi-bin/Tools/w3pimms/w3pimms.pl}.}. In both cases, we considered the energy range of $0.3-2.0$ keV and the thermal spectrum with the temperature of $100$ eV. For the NICER telescope, we adopt the minimal count rate of $0.51$ counts/s in the 0.3-2.0 keV band, which corresponds to $f_{X}\approx 4.3 \times 10^{-13}\,{\rm erg\,cm^{-2}\,s^{-1}}$ \citep{2022AJ....163..130R}. Assuming the characteristic X-ray luminosity of the eruptions to be $L_{X}=5\times 10^{42}\,{\rm erg\,s^{-1}}$, we estimate the maximum luminosity distance (cosmological volume) which the current detectors can probe using the simple relation, $d_{\rm L}=[L_{X}/(4\pi f_{X})]^{1/2}$. Thus, we get $d_{\rm L}\sim 814$ Mpc and $\sim 3613$ Mpc for the Swift and the XMM-Newton telescopes, respectively. For the NICER X-ray timing telescope, the limiting luminosity distance is $\sim 312$ Mpc. Overall, this large scatter in the cosmological volume justifies the distance cut-off of 1 Gpc. The caveat of the simple estimates above is the assumption that the EM counterparts of gas-rich EMRIs are all similar to the currently detected soft X-ray QPEs. In principle, some of them can be more or less luminous and could thus manifest in different wavebands, in particular harder X-ray or softer bands such as UV, which directly affects the number of detected sources. Below we discuss some of those aspects considering the current and future survey telescopes, including radio and UV surveys in addition to X-ray telescopes. 

\begin{figure*}
    \centering
    \includegraphics[width=\columnwidth]{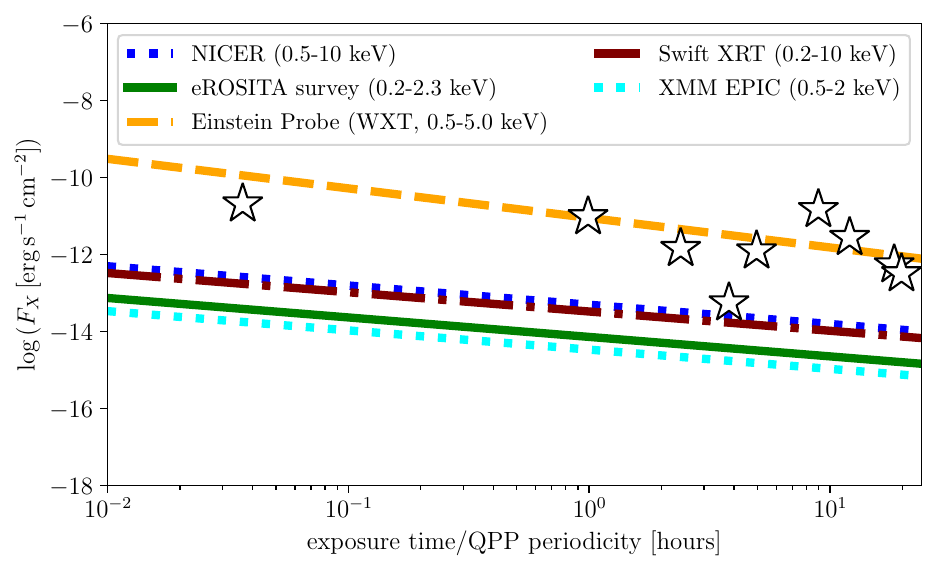}
    \includegraphics[width=\columnwidth]{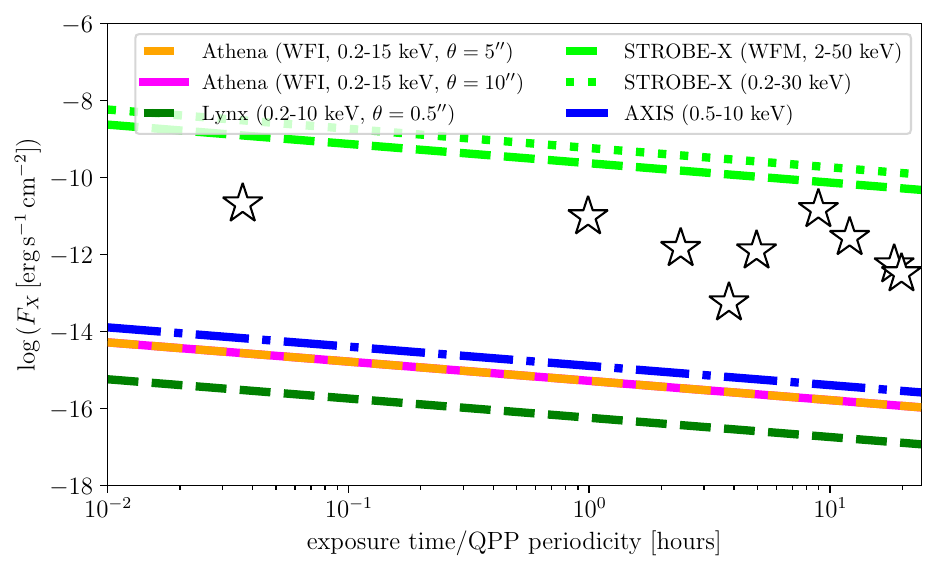}
    \caption{The flux density or flux sensitivity (in ${\rm erg\,\,s^{-1}\,cm^{-2}}$) as a function of the cumulative exposure time (in hours). \textit{Left:} Sensitivity levels as a function of exposure time (in hours) for currently used instruments capable to detect soft X-ray emission (NICER, Einstein Probe, Swift XRT, XMM EPIC, eROSITA). The stars represent already detected repeating X-ray transients with a peak flux density and periodicity (QPPs; see Table~\ref{tab:QPsummary}). \textit{Right:} Same as in the left panel but for some of the planned X-ray missions (Athena, Lynx, AXIS, STROBE-X). These levels should be considered as estimates since the actual sensitivity depends on the spectrum of the source, cosmic and galactic X-ray background, and the real-time performance of the detector in space.   }
    \label{fig_flux_time}
\end{figure*}
\subsubsection{Detectability by specific X-ray missions}

Currently detected soft X-ray repeating sources (QPPs) are plotted in Fig.~\ref{fig_flux_time} along with the flux sensitivity limits of current (left panel) and future-planned detectors in the soft X-ray band (right panel). From this, we infer that the NICER telescope could detect most currently observed sources since they are above the flux sensitivity limit \citep[blue dotted line in Fig.~\ref{fig_flux_time};][]{2022AJ....163..130R}, while the SRG/eROSITA survey in soft X-rays (0.2-2.3 keV) \citep[green solid line;][]{2021A&A...647A...1P} and Swift XRT (0.2-10 keV)~\citep[maroon dash-dotted line;][]{2005SSRv..120..165B} should be able to detect all of these sources. XMM-Newton EPIC detector (0.5-2 keV)~\citep[cyan dotted line;][]{2001A&A...365L..45H,2008A&A...479..283B} is the most sensitive in the X-ray domain from the list of current instruments. The Einstein probe launched on January 9, 2024 \footnote{See e.g. \url{https://heasarc.gsfc.nasa.gov/docs/objects/heapow/archive/technology/einsteinprobe.html}.}, will have a wide field of view ($\sim 3600\,{\rm deg^2}$) using the Lobster-eye technology (the Wide-field X-ray Telescope, WXT) and will be able to detect about half of the currently detected transients \citep[dashed orange line; ][]{2022hxga.book...86Y} at 0.5-4 keV. The future \textit{Athena} mission~\citep{2013arXiv1306.2307N}, currently under preparation, with a field of view of $0.4\,{\rm deg^2}$ (Wide Field Imager) and angular resolution of 5-10 arcseconds, will be able to detect even fainter sources in the soft X-ray band with the flux density at least two orders of magnitude lower (see magenta and orange lines in the right panel of Fig.~\ref{fig_flux_time}). The concept of the \textit{Lynx} mission \citep{2019JATIS...5b1001G} with a smaller field of view of 0.1${\rm deg^2}$ and an angular resolution of 0.5'' should reach even lower flux densities (dashed green line). \footnote{The \textit{Athena} and \textit{Lynx} sensitivity curves in Fig.~\ref{fig_flux_time} were calculated following the proposed X-ray observatories O1, O2, and O3 according to \citet{2023MNRAS.519.5962L}.} The concept of Advanced X-ray Imaging Satellite (AXIS) mission \citep[0.5-10 keV;][]{2019BAAS...51g.107M,2023arXiv231107658A} with the exquisite spatial resolution ($<1.5''$) across 24' field of view will reach the flux sensitivity intermediate between the planned Athena and the current XMM EPIC detector. The planned STROBE-X Wide Field Monitor as well as its X-ray Concentrator Array \citep[2-50 keV and 0.2-12 keV, respectively, ][]{2019arXiv190303035R} will reach the sensitivity to detect just the brightest QPP sources (dashed and dotted lime lines in Fig.~\ref{fig_flux_time}). The detection and classification of new X-ray QPP sources will benefit from the combination of current and future survey telescopes, such as eROSITA and Einstein probe, with the dedicated monitoring of the detected QPPs with more sensitive detectors, e.g. on-board of XMM-Newton and Athena. 
We note that Athena and Lynx missions are currently planned to be launched in the mid-2030s, hence they are not relevant for the detection of EMRI EM counterparts ``today'', though they will be crucial for such detections and monitoring in the era of future space-borne GW detectors, possibly including LISA. 

\subsubsection{Dependence on the accretion state and prospects for detectability in the UV domain}
The prospect of detecting EMRIs in the electromagnetic spectrum can also be a function of the relative accretion rate, as suggested by \citet{linial2023emri_UV}. Within their model where the QPEs are generated by the ejection of shocked plasmoids as the CO/star punches through the accretion disk, the flares can effectively be detected when their radiation is spectrally harder and hence brighter in the soft X-ray band than that of the underlying accretion flow. This is not met for accretion rates close to the Eddington limit, i.e. when the relative accretion rate (or equivalently the Eddington ratio) is $\dot{m} \equiv \dot{M}/\dot{M}_{\rm Edd} \sim 1$ \citep[where $\dot{M}_{\rm Edd}$ is the Eddington accretion rate as defined in][]{linial2023emri_UV}. As the relative accretion rate $\dot{m}$ drops to $\dot{m} \sim 0.1$, e.g. following a tidal disruption event with a decreasing accretion rate, QPE emission in the soft X-ray band is above the quiescent disc emission due to the higher temperature associated with QPEs. \citet{linial2023emri_UV} discuss that for a further drop in the accretion rate to $\dot{m} \sim 0.01$, the QPEs become too dim in the X-ray domain due to softer radiation. At this stage, the eruptions could rather be traceable in the UV domain. An example of such sources is provided by the AGN QPE source, GSN069, with its generally decreasing accretion rate following two TDE-like outbursts. Eruptions in the source were not detected for accretion rates $\dot{m}\geq 0.5$, but (re)appeared for lower accretion rates \citep{Miniutti2019,2023A&A...670A..93M,2023A&A...674L...1M}. 
Future monitoring of this source by X-ray and UV instruments will reveal how the eruption SED changes with the drop in the accretion rate. A drop in eruption amplitudes that can be potentially linked to a decrease in accretion rate is also traced for eRO-QPE1 \citep{Pasham_2024,2024arXiv240208722C}. Note that a UV-band luminosity of $\sim 10^{41} \rm erg s^{-1}$, typical for such sources, corresponds to the apparent AB magnitude of $\sim 21.8$ mag for nearby sources ($z \sim 0.025$). Such sensitivity will be reached by the future-planned wide field-of-view NUV satellite ULTRASAT \citep{2022SPIE12181E..05B,2023arXiv230414482S}, which will perform continuous monitoring of the transient UV sky (230-290 nm) thanks to its wide field-of-view of $204\,{\rm deg^2}$. It will also provide target-of-opportunity triggers for the planned two-band (FUV, 140-190 nm, and NUV, 260-360 nm) UV photometry mission QUVIK \citep{2023arXiv230615080W,2023arXiv230615081K,2023arXiv230615082Z}, which will be able to perform a high-cadence dedicated monitoring of UV QPP candidates down to $\sim 22$ and $\sim 20$ AB magnitudes in the NUV and FUV bands (detector SNR of 5 in 1000 seconds) with the point spread function of $\lesssim 2.5''$. In addition, the UVEX mission \citep{2021arXiv211115608K} with the planned launch in the early 2030s will perform medium-resolution spectroscopy in the range 115.0–265.0 nm in addition to deep two-band monitoring of the UV sky with its sensitivity reaching $\sim 24.5$ AB magnitude (detector SNR of 5 in 900 seconds).  Sensitive UV observations may also reveal long-period UV QPEs tracing EMRIs crossing the disc on wider orbits, i.e. at lower relative velocities producing softer thermalized shocks. Alternatively, short-period QPEs around low-mass central black holes of $\lesssim 10^{5.5}\,M_{\odot}$ are also expected to be detected in the UV domain \citep{linial2023emri_UV}.

\subsubsection{Prospects for coincident detection in the radio band}

The detection of gas-rich EMRI in the radio domain is highly uncertain, depending mainly on the presence of the magnetized outflow or jet in EMRI hosts, i.e. whether the source is radio-loud or quiet. The viewing angle also affects the jet contribution via the Doppler-boosting factor. One of the most promising sources for hosting a binary MBH -- OJ287 -- is a blazar that exhibits a quasiperiodic precession/nutation pattern in the temporal evolution of the jet position as well as its radio brightness on the timescale of $\sim 20$ years \citep{2018MNRAS.478.3199B,2023ApJ...951..106B}, which is about twice longer than the typical optical variability timescale in this source \citep{2016ApJ...819L..37V}. However, most of the QPP hosts in Fig.~\ref{tab:QPsummary} are radio quiet and hence the prospects for tracing the EMRI presence in the radio variability is at best at the limit of the sensitivity of current radio telescopes. From the repeating soft X-ray transients (QPE-like) only Swift J0230+28 exhibited one radio flare coincident with the start of the X-ray flare \citep{2023arXiv230903011G}. The radio flare detected by the Very Large Array (VLA; D configuration) was unresolved and coincident with the point source in the nucleus. The flux density at 10 GHz was $F_{\nu}=93 \pm 7 \mu {\rm Jy}$ (13$\sigma$ detection). Given the distance of the host galaxy ($\sim 165\,{\rm Mpc}$), the radio luminosity at 10 GHz is $L_{\rm 10GHz}\sim 3 \times 10^{37}\,{\rm erg\,s^{-1}}$ and the ratio of the radio luminosity to the corresponding soft X-ray luminosity of the flare is $L_{\rm 10GHz}/L_{X}=3 \times 10^{37}/(10^{42})\sim 3\times 10^{-5}$, hence the energy going into the radio emission is significantly smaller than the detected X-ray luminosity. The radio flare could be related to the synchrotron emission due to star-disc interaction shocks or the magnetic reconnection event in the magnetized plasma of the hot accretion flow. For the known QPPs, one can plan observational campaigns of simultaneous X-ray/radio monitoring with the VLA, whose sensitivity threshold is as low as $10\,{\rm \mu Jy}$ (RMS noise) at 10 GHz (bandwidth 500 MHz, field of view of 4.2 arcmin) for the on-source time of 24.9 minutes \footnote{Calculated using the on-source time calculator \url{https://obs.vla.nrao.edu/ect/}.}. In the second half of 2020s, the Square Kilometer Array (SKA-1, mid-frequency) should reach the RMS noise level of $1.2\,{\rm \mu Jy}$ for 1 hour-long observation at 12.5 GHz (bandwidth 3.75 GHz), which should reveal radio variability of even fainter QPPs. With the field of view of 6.7 arcmin (SKA1-mid, 12.5 GHz) and a high sensitivity, the SKA will survey the sky faster than previous radio arrays, making it possible to visit the sky more frequently \citep{2019arXiv191212699B}. For detecting radio flares from previously unidentified QPPs, hence candidates for EMRIs, a large sky coverage in the radio domain is desirable. The concept of the Canadian Hydrogen Intensity Mapping \citep[CHIME; ][]{2022ApJS..261...29C} experiment with the total sky coverage of 31\,000 ${\rm deg^2}$ for one-year integration is in principle beneficial for detecting new radio transients such as fast radio bursts. However, the confusion noise level of $\sim 0.1$ Jy currently limits the instrument to much brighter transients than those expected from CO-accretion flow interactions in nominally non-jetted hosts (at least thousand times brighter than the Swift J0230-like radio eruption).    

\subsection{QPP luminosity and dependence on binary parameters}\label{sec:caveats}
The dynamics of the putative binaries behind QPPs will depend on a number of their properties, such as the mass of the perturbing COs, the properties of the accretion disk, or the orbital parameters. Here we discuss how these will affect the character and observability of the QPPs.

\subsubsection{Radius of disk-perturber interaction and QPE luminosity}
\label{sssec:radlum}
Using the results on the orbital frequency of EMRI detectable EM candidates (see Fig.~\ref{fig:JointObservation}) we set the fiducial orbital period to $P_{\rm orb}=1$ hour (QPE-like eruption occurs every $\sim 0.5$ hours for the CO-disc interaction). The primary MBH mass is set to $M=10^6\,M_{\odot}$ unless stated otherwise. Then the radius of the circular orbit of the CO is 
\begin{align}
r=\left( \frac{P c^3}{GM}\right)^{2/3}\,r_{\rm g} \sim 24 \left(\frac{M}{10^6 M_\odot}\right)^{-2/3} \left(\frac{P}{1 \, \rm hour}\right)^{1/3} r_{\rm g}\,,
\end{align}
where $r_{\rm g} \equiv GM/c^2 \sim 1.5 \,{\rm km} (M/M_\odot) $ is the gravitational radius of the MBH. 

The influence radius $R_{\rm inf}$, within which the compact object or the perturber with the mass $\mu$ affects the surrounding gas gravitationally, depends on the relative velocity with respect to the accretion disc material $v_{\rm rel}$ and the disc sound speed $c_{\rm s}$ at the distance $r$, \citep[e.g.][]{Sukov__2021}
\begin{equation}
    R_{\rm inf}=\frac{2 G \mu}{v_{\rm rel}^2+c_{\rm s}^2}\,.
    \label{eq_influence_radius}
\end{equation}
Approximating both the disc velocity and the perturber velocity with Keplerian radial profiles, and adopting the inclination $\iota$ between the perturber and the accretion disc, we may estimate the relative velocity as follows,
\begin{equation}\label{eq:relativevelocity}
    v_{\rm rel}=\left[\frac{2GM}{r}(1-\cos{\iota}) \right]^{1/2}\,.
\end{equation}
The sound speed is calculated as $c_{\rm s}=(k_{\rm B}T_{\rm disc}/m_{\rm p})^{1/2}$, where $k_{\rm B}$ is the Boltzmann constant, $T_{\rm disc}$ is the temperature of the assumed standard disk, and $m_{\rm p}$ is the proton mass. 

Specifically, we can expect the accretion disk sonic speed to be $c_{\rm s} \lesssim v_{\rm k} \sim \sqrt{G M/r}$, where $v_{\rm k}$ is the Keplerian speed at radius $r$. For a radiatively efficient accretion disk we will have $c_{\rm s}/v_{\rm k} \ll 1$ and the height of the disk $H/r \sim c_{\rm s}/v_{\rm k}  \ll 1$ \citep{ShakuraSunyaev1973}, whereas for hot radiatively inefficient flows we will have $H/r \sim c_{\rm s}/v_{\rm k} $ of order one \citep{2014ARA&A..52..529Y}. We thus obtain $R_{\rm inf}$ as
\begin{align}
    R_{\rm inf} = r \frac{\mu}{M} \left( 1 - \cos \iota + \frac{1}{2}\left(\frac{H}{r}\right)^2 \right)^{-1} \label{eq_Rinf}
\end{align}

This estimate for $R_{\rm inf}$ ignores the gravitational influence of the primary MBH, which will dominate the gas outside the distance given by the Hill (tidal) radius $R_{\rm Hill}$ from the CO, where
\begin{equation}
    R_{\rm Hill}\sim r \left(\frac{\mu}{3 M} \right)^{1/3} \,.
    \label{eq_Hill_radius}
\end{equation}
 It is easy to see that for moderate inclinations and small mass ratios $\mu/M$, the Hill radius will always be larger than $R_{\rm inf}$, so eq. \eqref{eq_Rinf} provides a good estimate for the interaction cross-section. For low inclinations $\sin \iota \lesssim (\mu/M)^{1/3}$ and $(H/r) \lesssim (\mu/M)^{1/3}$, however, the Hill radius acts as a cap for the radius affected by the perturber.

Additionally, we conjecture that a QPE will only be created by an orbiter that is both not embedded in the disk, and punches through the disk at a relative speed that is supersonic, both of which lead to the same condition $\sin \iota \gtrsim H/r$.\footnote{Assuming perturbers co-rotating with the gas direction. There is an edge case of counter-rotating orbits (formally $\iota \sim 180^\circ$) which are embedded in the disk and sweeping through it at highly supersonic speeds.} For $\sin \iota \lesssim H/r$ the orbiter is embedded in the disk and we surmise that this instead leads to a QPO.

For example, at the distance of the CO corresponding to one hour orbital period, the sound speed of the accretion-disk gas is $\sim 35\,{\rm km\,s^{-1}}$ for a fiducial relative accretion rate of $\dot{m}=0.1$ and an innermost stable circular orbit set to $6\,r_{\rm g}$, while the Keplerian velocity is $\sim 61\,  000 \,\rm km\,s^{-1}$. As a result, the orbit will be non-embedded in the disk and cross it supersonically for $\iota \gtrsim 0.03^\circ$. Similarly, the orbit will not reach the cap on the influence radius set by the Hill radius for $\iota \gtrsim 0.5^\circ (\mu/M_\odot)^{1/3} (M/10^6 M_\odot)^{-1/3}$.

\begin{figure}
    \centering
    \includegraphics[width=\columnwidth]{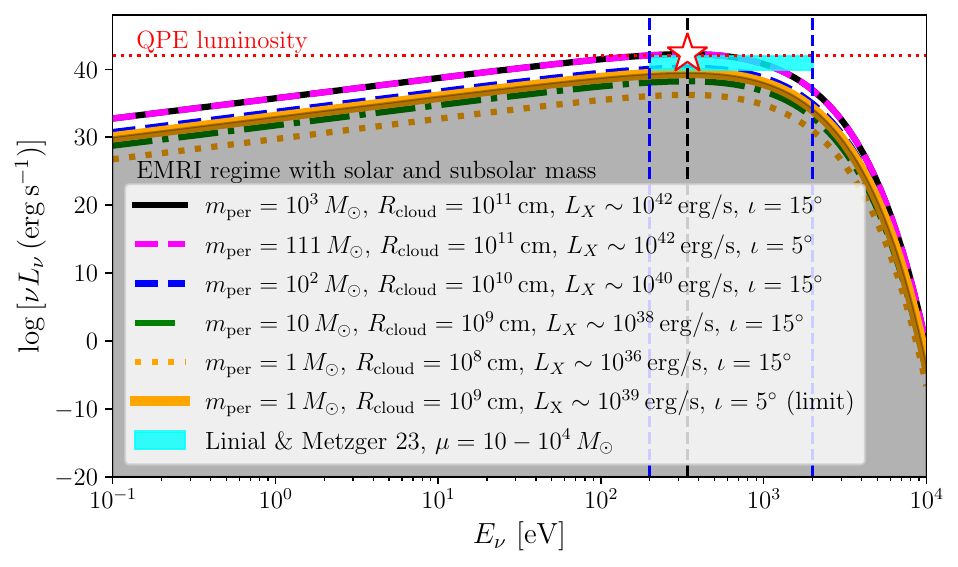}
    \caption{Exemplary spectral energy distributions (SEDs) corresponding to shocked clouds of material ejected during the passages of a lighter compact object (CO) around the MBH. We assumed the orbital period of 1 hour and the MBH mass of $10^6\,M_{\odot}$. To reach typical QPE luminosities of $10^{42}\,{\rm erg\,s^{-1}}$ (red dotted line) in the 0.2-2 keV band (blue shaded region), the initial cloud size should be $\sim 10^{11}\,{\rm cm}$, which corresponds to $\sim 0.7$ gravitational radii of the MBH. The compact objects that are able to eject such clouds fall into the range of $\sim 100-1000\,M_{\odot}$, with the inclination of $5-15$ degrees with respect to the accretion disc, respectively. Solar and subsolar COs typically generate X-ray flares of $\lesssim 10^{39}\,{\rm erg\,s^{-1}}$ (grey shaded region), hence at least three orders of magnitude below the typical QPE luminosities. The dashed black vertical line represents the peak of the SEDs. The white star stands for the mean QPE luminosity inferred from seven detected QPE sources ($\log{(L_{\rm x}\,[\rm erg\,s^{-1}])}=42.3 \pm 0.6$; see Table~\ref{tab:QPsummary}). The cyan rectangle represents the X-ray luminosity range for COs of $\mu=10-10^4\,M_{\odot}$ according to the CO-disk interaction model by \citet{linial2023emri}.}
    \label{fig_Xray_lum_cloud}
\end{figure}

To estimate the X-ray luminosity of QPE-like eruptions, we then adopt a toy model of CO-disc interactions, which lead to shocks and ejection of shocked spherical, optically thick clouds above the disc plane \citep[see e.g.][]{Franchini_2023}. For the initial cloud radius, we adopt $R_{\rm inf}$ and for its temperature, we set $T_{\rm cloud}\sim 10^6\,{\rm K}$, which was inferred for QPEs \citep{Miniutti2019,Arcodia2021,2024arXiv240117275A}. Then the soft X-ray luminosity (in the 0.2-2 keV band) can be estimated as follows,
\begin{equation}
    L_{\rm X}=4 \pi R_{\rm inf}^2 \int_{0.2\,{\rm keV}}^{2.0\,{\rm keV}} \frac{2 h \nu^3}{c^2} \frac{\mathrm{d}\nu}{\exp{(h\nu/k_{\rm B}T_{\rm cloud})-1}}\,,
    \label{eq_xray_luminosity}
\end{equation}
where $\nu$ is the frequency corresponding to the photon energy $E_{\nu}=h\nu$ and $h$ is the Planck constant. We plot examples of the spectral energy densities corresponding to this model in Figure \ref{fig_Xray_lum_cloud}.

\citet{linial2023emri} present a similar CO-disk interaction model, taking into account inefficient photon production during the optically thick phase of an expanding shocked cloud. To reach $L_{X}\sim 10^{42}\,{\rm erg\,s^{-1}}$, the CO mass should fall into the IMBH range with $\mu\sim 10^4\,M_{\odot}$,
\begin{align}
\begin{split}
    L_{\rm X} &= 8.9 \times 10^{41} \left(\frac{\dot{m}}{0.1} \right)^{1/3}  \left(\frac{M}{10^6\,M_{\odot}} \right)^{5/9} \,\\
    & \times \left(\frac{\mu}{10^4\,M_{\odot}} \right)^{2/3}  \left(\frac{P_{\rm QPP}}{0.5\,\text{hours}} \right)^{-2/9}\,{\rm erg\,s^{-1}}.
    \label{eq_linial_metzger_QPE}
\end{split}
\end{align}
We indicate the luminosity range according to Eq.~\eqref{eq_linial_metzger_QPE} in Figure~\ref{fig_Xray_lum_cloud} using the cyan rectangle, which corresponds to the CO masses of $\mu=10-10^4\,M_{\odot}$. 

\subsubsection{Theoretical uncertainties of disk-perturber interaction}
\label{sssec:uncert}
The \citet{ShakuraSunyaev1973} $\alpha$-disk used for various estimates above corresponds only to the effective structure of an accretion disk averaged over the orbital period, disk height, and azimuthal angle. As such, it cannot be used for the dynamical analysis of its interactions with bodies embedded in the disk (i.e., smaller than the disk height) and/or moving on the orbital time-scale, since on those scales the structure is dominated by magneto-rotational-instability-driven turbulence (see \citet{1991ApJ...376..214B} and especially the discussion in \citet{1998RvMP...70....1B}). 

Another difficult aspect of the interaction are the dynamics of the ejected plasma leading to the observed lightcurve of the QPE. In some models the ejected material is assumed to expand in the form of a ``bubble'' for some time before becoming optically thin and releasing the radiation \citep{LehtoValtonen1996,Ivanov1998,2016MNRAS.457.1145P,linial2023emri}. However, the expansion implies significant adiabatic cooling of the plasma, which requires involving other subtle mechanisms to explain the relatively high $10^6 \, \rm K$ effective temperatures observed in QPEs \citep{linial2023emri}. Additionally, the CO will eject the material in the form of a trailing wake as seen in the Hoyle-Lyttleton scenario, which is likely incompatible with the expanding bubble model. Furthermore, for some configurations, such as lower-inclination and retrograde orbits, the break-out emission of a bow shock associated with the supersonic CO can dominate the EM signal \citep{2023MNRAS.526...69T}. As such, both the action of the perturber on the disk, and the possible back-reaction of the disk on the perturber is currently uncertain and will likely be only resolved by first-principle simulations of the interaction of the plasma of the accretion disk with the orbiting body.

Furthermore, the model of star-disk interactions described above and the estimated X-ray luminosity rely on the standard geometrically thin and optically thick accretion disc. However, most of the EMRI hosts in the local Universe are expected to be low-luminosity sources. Such sources host hot accretion flows that are advection-dominated and radiatively inefficient \citep{2014ARA&A..52..529Y}. In this case, plasma is more diluted and warmer, hence the shocks due to interactions with the CO are also weaker because of lower Mach numbers. The ejected mass and the associated electromagnetic energy is also lower due to a smaller surface density \citep{linial2023emri}. For the hot flow, the variability could rather be driven by the accretion-rate variations due to propagating density waves induced by the CO bow shock \citep[see e.g.][]{Sukov__2021}. 

\subsubsection{Mass of compact object and detectability}
\label{sssec:co_mass}
 The mass range of orbiting COs in EMRIs/IMRIs is rather broad -- from a few Solar masses \citep[neutron stars, white dwarfs, stellar black holes;][]{1983bhwd.book.....S} to several thousand or even ten thousand in the case of intermediate-mass black holes \citep[IMBHs; ][]{2020ARA&A..58..257G,2024arXiv240210140P}. The range can be extended to lower, sub-solar masses if we admit the occurrence of brown dwarfs in galactic nuclei \citep[see, e.g.,][]{2019PhRvD..99l3025A,2019A&A...627A..92G,2021ApJ...915..111Z} or hypothetical light primordial black holes \citep[see, e.g.,][]{2022PhRvL.129f1104A,2023PDU....4101231B}.
 
In Fig.~\ref{fig_Xray_lum_cloud}, we plotted spectral energy distributions (SEDs) of emitting shocked clouds ejected from the accretion disc due to the passages of lighter compact objects. To reach typical QPE X-ray luminosities of $L_{X}\sim 10^{42}\,{\rm erg\,s^{-1}}$ (0.2-2.0 keV; red dotted line), the initial cloud size should be $R_{\rm in}\sim 10^{11}\,{\rm cm}$. This corresponds to the influence radius of a CO with mass $\mu \sim 10^2-10^3\,M_{\odot}$ 
orbiting around the MBH at the inclination of $5-15$ degrees with respect to the accretion disc (prograde sense), respectively. This range corresponds to rather low inclinations. If the mutual inclination is increased so that the CO and the disc are approximately perpendicular to each other, then the required CO mass is in the IMBH range, $\mu \sim 29\,000\,M_{\odot}$, since the relative velocity increases to $v_{\rm rel}\sim 86\,400\,{\rm km\,s^{-1}}$. If we keep the same inclination of 15 degrees and decrease the perturber mass by one order of magnitude, the influence radius decreases by one order and the X-ray luminosity by two orders of magnitude, see Eqs.~\eqref{eq_influence_radius}, \eqref{eq_xray_luminosity}, and Fig.~\ref{fig_Xray_lum_cloud}. 

For Solar-mass COs, the X-ray luminosity is typically $L_{X}\lesssim 10^{39}\,{\rm erg\,s^{-1}}$ (solid orange line in Fig.~\ref{fig_Xray_lum_cloud}) for the MBH mass of $\gtrsim 10^5\,M_{\odot}$ and inclination of $\gtrsim 5$ degrees, which corresponds to the relative velocity of $\gtrsim 2500\,{\rm km\,s^{-1}}$ (the orbital distance is $111\,r_{\rm g}$ for $M=10^5\,M_{\odot}$, one hour orbital period). For the minimal inclination of $\iota\sim 0.5^{\circ}$ around $M=10^6\,M_{\odot}$, the X-ray luminosity could reach $L_{X}\sim 10^{42}\,{\rm erg\,s^{-1}}$ assuming $T_{\rm cloud}\sim 10^6\,{\rm K}$, which is, however, a special ``grazing'' setup. Using the CO-disk interaction model of \citet{linial2023emri}, the luminosity estimate is $L_{X}\sim 2 \times 10^{39}\,{\rm erg\,s^{-1}}$ for $M=10^6\,M_{\odot}$ and $\dot{m}=0.1$. For the interactions of a Solar-type star with the disk, the same model predicts $L_{\star}\sim 1.4\times 10^{42}\,{\rm erg\,s^{-1}}$. However, Solar-type stars are tidally disrupted at $r\sim 47\,r_{\rm g}$, hence they are not relevant for QPPs with $f_{\rm QPP}\sim 0.5$ mHz. On the other hand, brown dwarfs with the mass of $\sim 0.08\,M_{\odot}$ and the radius of $\sim 0.1\,R_{\odot}$ would disrupt at $\sim r_{\rm t}/r_{\rm g}\sim 11\,(R_{\star}/0.1R_{\odot}) (M/10^6\,M_{\odot})^{-2/3} (\mu/0.08\,M_{\odot})^{-1/3}$, hence the brown dwarf-disk interactions with $f_{\rm QPP}\sim 0.5$ mHz are plausible. The X-ray flares would have luminosities one order of magnitude below the QPE luminosities \citep{linial2023emri},
\begin{align}
\begin{split}
    L_{\rm dwarf} &\sim 3\times 10^{41} \left(\frac{R_{\star}}{0.1\,R_{\odot}} \right)^{2/3} \left(\frac{M}{10^6\,M_{\odot}} \right) \,\\
    &\times \left(\frac{\dot{m}}{0.1} \right)^{1/3} \left(\frac{P_{\rm QPP}}{0.5\,\text{hours}} \right)^{-2/3} \,{\rm erg\,s^{-1}}.
\end{split}
\end{align}

Hence, the prospects for detecting X-ray flares from orbiting and inclined solar-mass and subsolar-mass COs are currently weak, but they can improve with the new generation of sensitive X-ray imaging instruments, such as Athena and Lynx. For instance, for the source at the redshift of $z=0.05$, the eruption X-ray flux density for the Solar-mass CO is predicted to be $F_{\rm X}\sim 10^{39}\,{\rm erg\,s^{-1}}/[4 \pi (222.3\,{\rm Mpc})^2]\sim 1.7 \times 10^{-16}\,{\rm erg\,s^{-1}\,cm^{-2}}$, which is at the sensitivity limit of Athena and Lynx for one hour exposure time, see Fig.~\ref{fig_flux_time}. Solar and subsolar-mass COs have characteristic influence radii smaller than one gravitational radius for typical parameters ($M=10^6\,M_{\odot}$). Using Eq.~\eqref{eq_influence_radius}, we obtain $R_{\rm inf}\sim 7 \times 10^{-4}\,r_{\rm g}$ for $\iota=15^{\circ}$, and using Eq.~\eqref{eq_Hill_radius}, we get $R_{\rm Hill}\sim 0.2\,r_{\rm g}$. Hence, this limits the overall strength of the accretion-disc perturbation and the resulting emission modulation.

We can also make rough estimates for CO masses causing significant perturbations of geometrically thick accretion disks based on the simulations of \citet{Sukov__2021}. In that paper it was found that in order to detect significant quasiperiodic variability in the inflow (accretion) rate, the size of the influence radius of the CO should be of the order of $R_{\rm inf}\sim 0.1-1.0\,r_{\rm g}$ for the CO distance of $\sim 10\,r_{\rm g}$. For $M=10^6\,M_{\odot}$, one gravitational radius we obtain the required CO mass in the range $\mu\sim 160-42\,000\,M_{\odot}$ (inclination between 5 to 90 degrees with respect to the disc midplane) to induce significant quasiperiodic variability. We can also compare this with the estimate using the Hill radius in eq.~\eqref{eq_Hill_radius}; the mass of the CO with the Hill radius equal to $1\, r_{\rm g}$ is
\begin{align}
\begin{split}
    \mu &=\frac{M^3}{P_{\rm orb}^2}\frac{12 \pi^2 G^2}{c^6}\,,\\
    &\sim 221\,\left(\frac{M}{10^6\,M_{\odot}} \right)^3 \left(\frac{P_{\rm orb}}{1\,\text{hour}} \right)^{-2}\,M_{\odot}\,,
\end{split}
\end{align}
which is within the same order of magnitude as the estimate using $R_{\rm inf}$.

\begin{figure*}
    \includegraphics[width=0.48\textwidth]{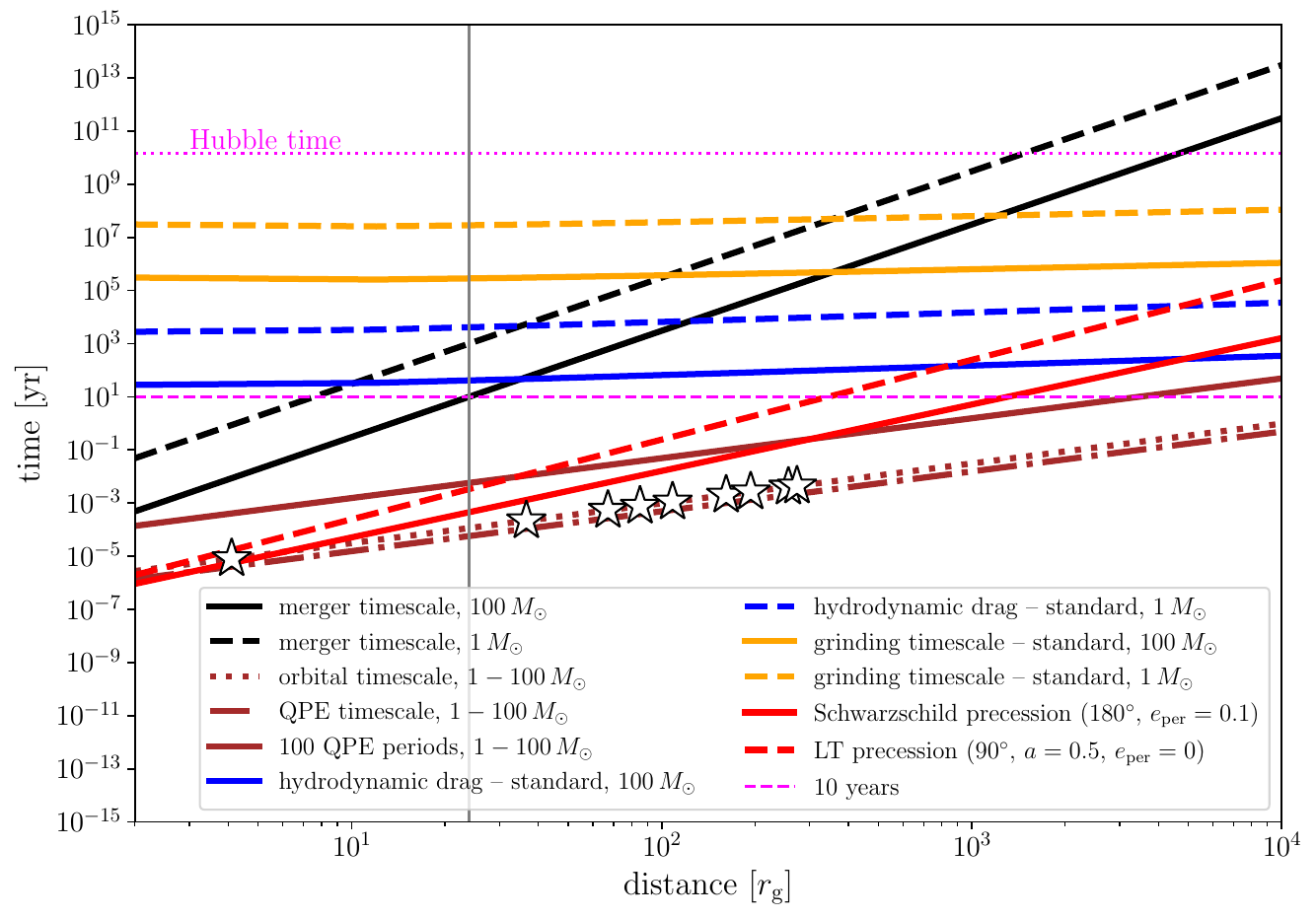}
    \includegraphics[width=0.48\textwidth]{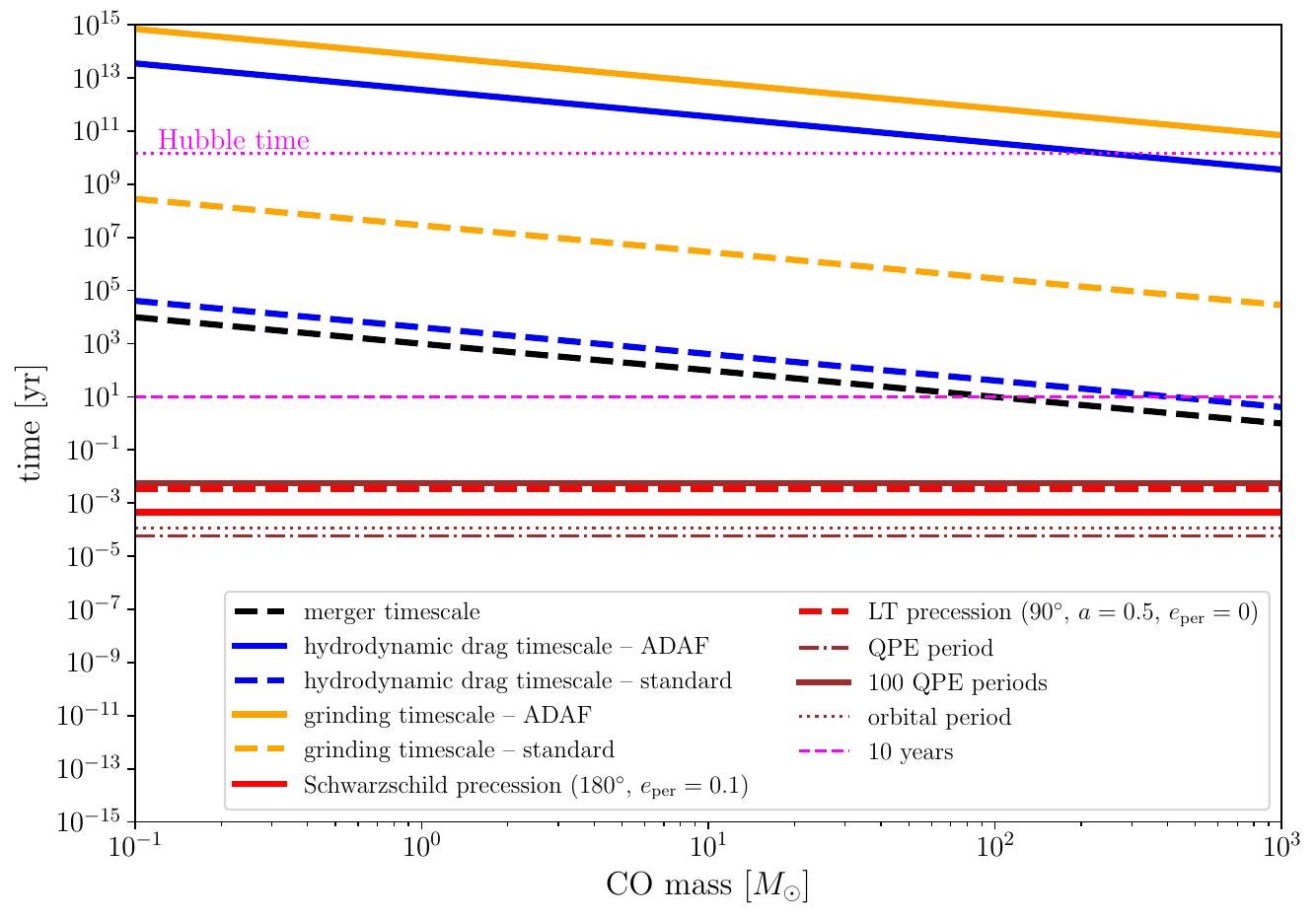}
    \caption{Comparison of different dynamical timescales for COs of different masses. \textit{Left:} Timescales expressed in years as a function of the distance from the MBH of $10^6\,M_{\odot}$ (in gravitational radii). Different dynamical processes (merger, grinding, hydrodynamical drag) are listed in the legend. Solid lines correspond to the CO mass of $100\,M_{\odot}$, while dashed lines correspond to the CO mass of $1\,M_{\odot}$. The vertical solid line at the distance of $\sim 24\,r_{\rm g}$ corresponds to the orbital period of one hour. The hydrodynamical and griding timescales are estimated for the standard disk with the relative accretion rate of $\dot{m}=0.1$ and the initial inclination of the CO set to $15^{\circ}$. The stars represent the positions of detected QPP sources (see Table~\ref{tab:QPsummary}) assuming that the CO orbital period is twice as long as the measured QPP period. The distance of the CO is estimated assuming a circular orbit around a $10^6\,M_{\odot}$ MBH. \textit{Right:} Timescales (in years) as a function of the CO mass (in Solar masses). The distance from the MBH of $10^6\,M_{\odot}$ is fixed to $\sim 24\,r_{\rm g}$, which corresponds to the orbital period of one hour. In this panel, we also depict separately the hydrodynamical-drag and grinding timescales for standard and hot accretion-flow discs adopting the relative accretion rates of 0.1 and 0.01, respectively.}
    \label{eq_timescales_CO}
\end{figure*}

Another issue with solar-mass and subsolar-mass compact objects is that they will not merge with the MBH within $\sim 10$ years, even for distances of the order of 10 gravitational radii. In Fig.~\ref{eq_timescales_CO} (left panel), we compare the merger timescale using the evolution model of \cite{1963PhRv..131..435P} for a CO of $100\,M_{\odot}$ (solid black line) with the one corresponding to $1\,M_{\odot}$ CO (dashed black line). For the CO of $100\,M_{\odot}$, the merger timescale is $\sim 10$ years for the mean distance of $\sim 24 r_{\rm g}$. We compare the merger timescales with the hydrodynamical grinding (or alignment) timescales \citep[][assuming the standard disk with $\dot{m}=0.1$ and the initial inclination of 15 degrees]{SyerClarkeRess1991} as well as with the hydrodynamical drag timescale related to dynamical friction due to the accretion-disc material \citep{1999ApJ...513..252O,2000ApJ...536..663N}, assuming the limiting case when the CO is embedded within the disk. We can notice that for heavier COs hydrodynamical timescales (alignment and drag) are shorter by a factor 100. By the same factor, the merger timescale is also shorter. The interaction of the embedded CO with the realistic accretion disk is expected to be more complex than the inspiral motion given by the classical hydrodynamic drag. Density waves can lead to both inward and outward type I migration and eventually to the formation of ``migration'' traps \citep{2010MNRAS.401.1950P,2016ApJ...819L..17B,Metzger_2022}. However, these dynamical effects are beyond the scope of the comparison of order-of-magnitude estimates provided here.  

The inspiral due to gravitational-wave emission dominates inside $\sim 35\,r_{\rm g}$ (merger timescales is shorter than the hydrodrag or dynamical friction timescale), while the merger timescale is shorter than the characteristic grinding timescale inside $\sim 350\,r_{\rm g}$. This indicates that inside $\sim 35\,r_{\rm g}$, where the EM counterparts of LISA EMRI sources are expected to be located, the gravitational-wave emission should be the main driver of the inspiral, with the dynamical friction force being progressively negligible (the associated timescale is longer than the merger timescale). At larger radii, the dynamical friction force due to the accretion-disc material as well as the hydrodynamical drag causing the alignment are expected to play dominant roles in bringing lighter perturbing objects into the closer vicinity of the MBH and aligning them with the accretion-flow plane.

In the right panel of  Fig.~\ref{eq_timescales_CO}, we fix the distance of the CO to $\sim 24\,r_{\rm g}$, which corresponds to one-hour orbital period for $10^6\,M_{\odot}$ MBH, and we show the dependency between the characteristic timescales and the CO mass. We see that the merger timescale is $\sim 10$ years for $\sim 100\,M_{\odot}$, while heavier COs would merge too soon (before the LISA launch), while lighter, solar-mass COs would merge on the timescale of $\sim 1000$ years. Similarly, it should be noted that the peak gravitational-wave strain of EMRIs scales linearly with $\mu$, so the GW detection distance of light COs will be smaller as compared to heavy COs even if they were close to merger within the LISA window. Combining the favourable merger timescale for the LISA observing window with the strong enough electromagnetic signal (see Subsec.~\ref{sssec:radlum}) gives preference for the CO mass of the order of $\sim 100\, M_{\odot}$, hence at the border between the stellar-mass and the intermediate-mass black holes. 

The origin and the dynamics of IMBHs in galactic nuclei are quite different from stellar-mass COs and the analysis of these differences is beyond the scope of the current manuscript. There may in principle also be new telltale features in the EM signature for higher-mass secondaries. For example, in the case of an IMBH orbiting in the plane of the accretion disk, one generically expects the perturbation to open a gap in the disk \citep{Syer:1995hk,2023MNRAS.522.2869S}. This gap later closes due to increasing radiation pressure at lower radii when the system is approaching the LISA frequency window, and the closure process leads to increased EM activity preceding the inspiral detection \citep{2011PhRvD..84b4032K}. However, \citet{2011PhRvD..84b4032K} estimate that for a standard $\alpha$-disk the gap is open only above the radius 
\begin{align}
    \frac{r_{\rm gap}}{r_{\rm g}} = 
    {\rm max} \left\{
        \begin{array}{l}
             20 \left(\frac{\alpha}{0.01}\right)^{1/5} \frac{\dot{m}}{0.1} \left(\frac{M}{10^5 M_\odot}\right)^{2/5}\left(\frac{\mu}{100 M_\odot}\right)^{-2/5} \\
             30 \left(\frac{\alpha}{0.01}\right)^{1/2} \frac{\dot{m}}{0.1} \left(\frac{M}{10^5 M_\odot}\right)^{1/2}\left(\frac{\mu}{100 M_\odot}\right)^{-1/2} 
        \end{array}
    \right\}\,.
\end{align}
At higher secondary masses, this transition occurs closer to the primary BH and thus makes it more relevant for our discussion. For values $\mu \sim 100 M_\odot $ and $M \sim 10^5 M_\odot$,\, $\alpha \sim 0.01$ and accretion at $10\%$ of the Eddington rate the gap would close below $r_{\rm gap} \sim 30 r_{\rm g}$ or for orbital periods above $\sim 500$ seconds. This implies that the closure of the gap occurs very close to the periods of hours to minutes of the QPPs we consider here. The associated increase of EM activity thus represents a new possible avenue for detecting and characterizing these systems. 

\subsubsection{Eccentricity and disk precession}
\label{sec:eccinc}

\begin{figure}
\centering
\includegraphics[width=0.48\textwidth]{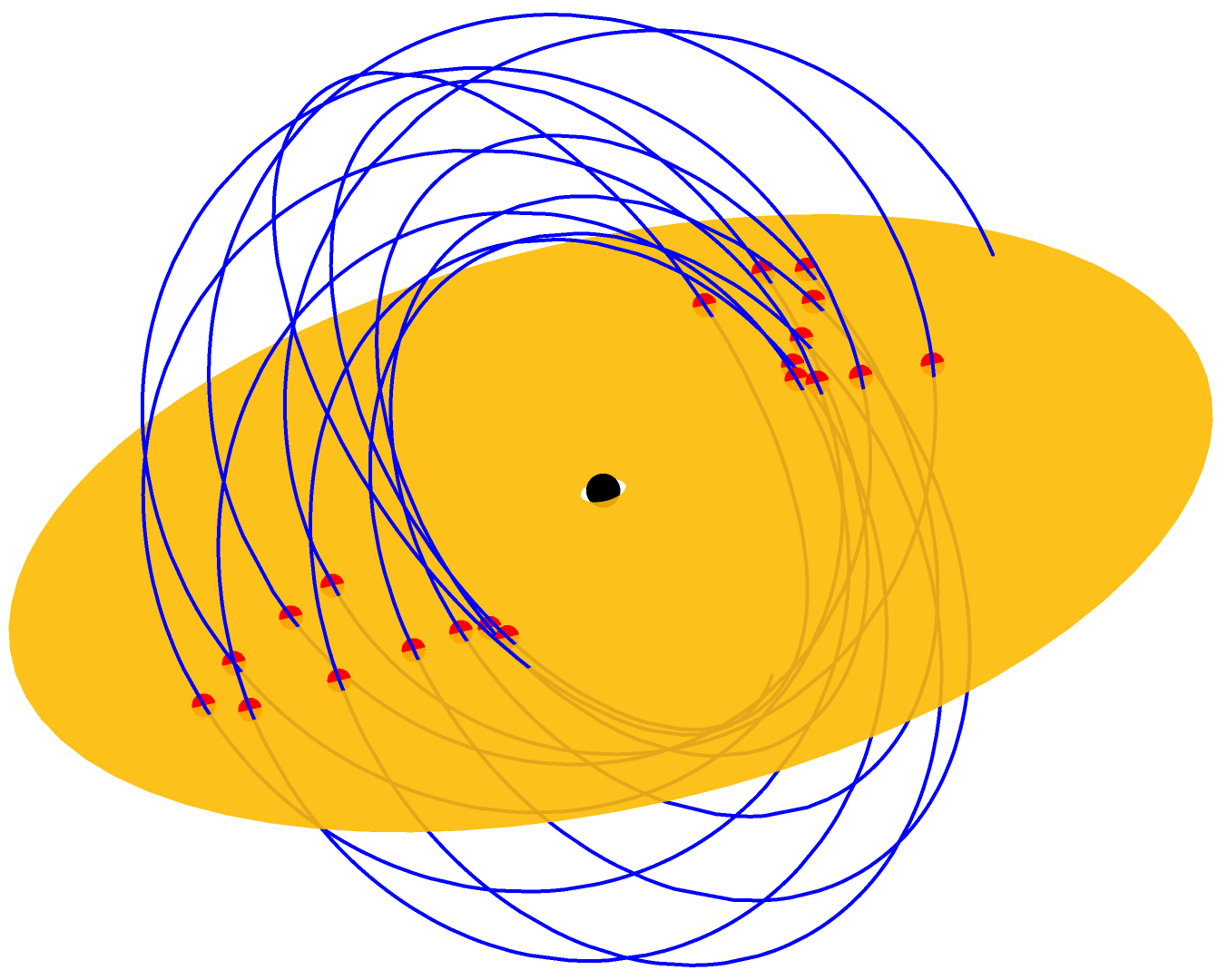}
\caption{The intersections (red) of an eccentric and inclined free test particle orbit (blue) through an accretion disk (orange) lying in the equatorial plane of a Kerr black hole with spin $a=0.98$ (black). The orbit is prograde with the orbital parameters semi-latus rectum $p=24 r_{\rm g}$, eccentricity $e=0.3$ and inclination $i = 60^\circ$ \citep[defined as in][]{schmidt2002celestial}. The interplay of eccentricity, pericenter precession, Lense-Thirring precession, and other sub-leading relativistic effects lead to a complex pattern.} 
\label{fig:orbit}
\end{figure}

The \texttt{M1} EMRI catalogue from \citet{Babak_2017} used in Figs \ref{fig:evolution}, \ref{fig:initdistribution} and \ref{fig:JointObservation} corresponds to the MBH mass function of \citet{Barausse:2012fy}, with the assumption that typical MBHs are rapidly spinning, $a_{\rm MBH} \sim 0.98 M$, typical mass of the secondary CO $\sim 10 M_\odot$, and specific assumptions about the cusp erosion, regrowth, and structure of the star cluster in the MBH environments \citep[see][ for more details]{Babak_2017}. In particular, the \texttt{M1} model assumes that the dominant formation channel for EMRIs is the already mentioned dynamical capture in a ``loss-cone'' scenario \citep{HilsBender1995,SigurdssonRees1997}. This leads to inspirals that have eccentricities $e\gtrsim 0.1$ during the entire course of their inspiral as can be seen from Figs. \ref{fig:evolution} and \ref{fig:initdistribution}, and only weak correlation of the orbital plane with the direction of the central MBH rotation. In other words, the EMRIs in question have generic inclinations and eccentricities as illustrated in Fig. \ref{fig:orbit}. On the other hand, the other already mentioned channels including the capture and separation of a binary by the Hills mechanism \citep{2005ApJ...631L.117M}, and disk-assisted migration \citep{Pan_2021} lead to low-eccentricity EMRIs with low inclinations in the latter cases and generic inclinations in the former. Hence, it is important to discuss what these distributions of orbital parameter imply for EM observations. Here we show in particular that generic orbits will lead to subtle timing features that will make the matching of QPPs to EMRIs nontrivial. 

\begin{figure}
\centering
\includegraphics[width=0.47\textwidth]{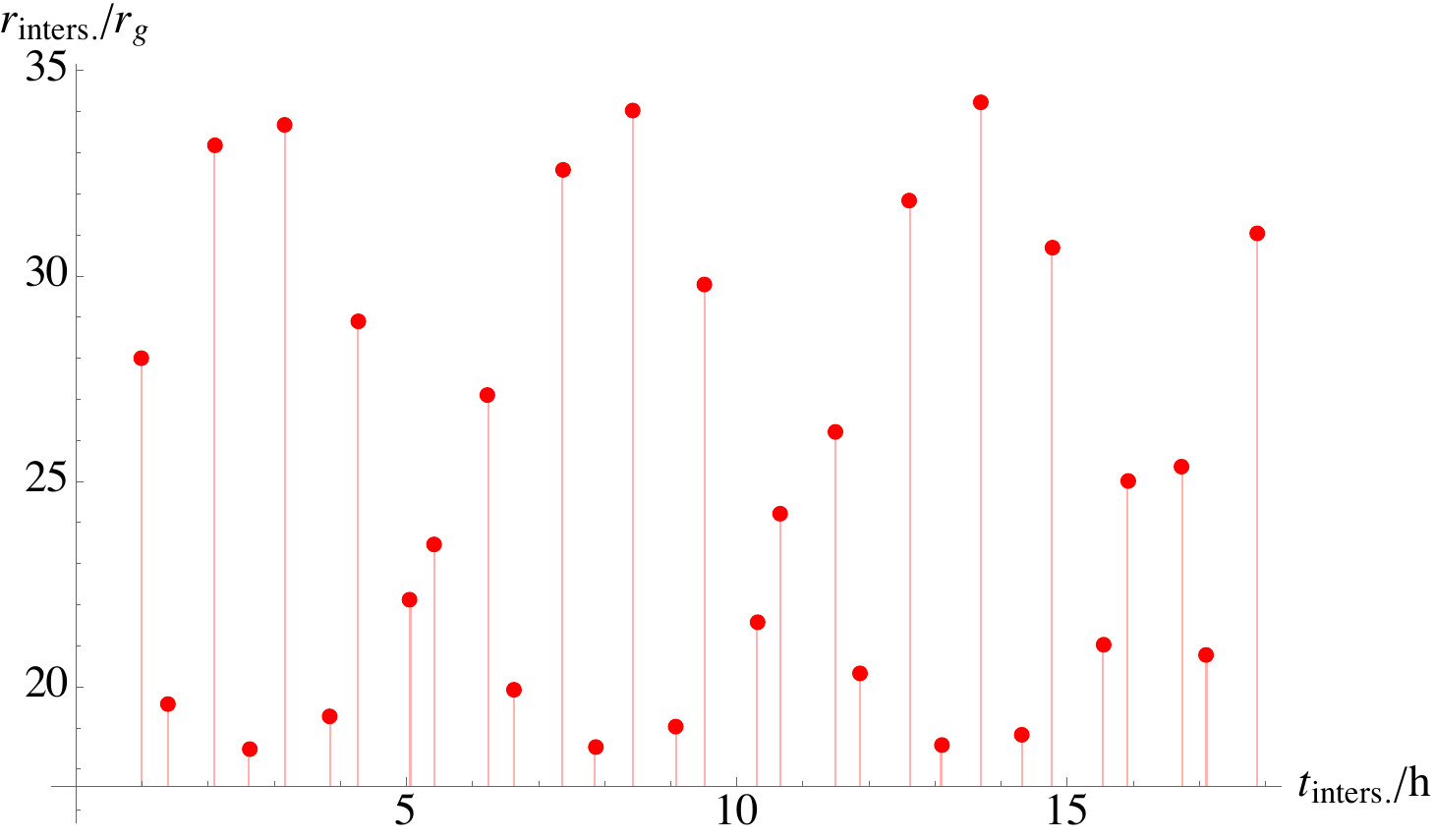}
\caption{30 intersection times and radii of the orbit from  Fig. \ref{fig:orbit} where the times correspond to the fiducial MBH mass $M=10^6 M_\odot$.}
\label{fig:interstime}
\end{figure}

If the perturbing CO evolves along an eccentric trajectory, this introduces a more complex two-periodic structure of the time-variability. With the leading time-scale being the azimuthal orbital period $\Omega_{\phi}\sim 2\pi/P_{\rm orb}$, the second time-scale is the pericentre precession frequency, which reads to leading post-Newtonian and mass-ratio order \citep{robertson1938note}
\begin{equation}
    \Omega_{\rm p} = \frac{3 (2 G M \pi)^{2/3}}{P_{\rm orb}^{5/3} c^2(1-e^2)} \sim \frac{0.13}{\rm hour} \frac{1}{1 - e^2}  \left(\frac{M}{10^6\,M_{\odot}} \right)^{2/3} \left(\frac{P_{\rm orb}}{1\,\text{hour}} \right)^{-5/3} \!,
\end{equation}
where $e$ is the eccentricity of the orbit. 
Specifically, if one assumes that the variability is caused by the CO punching through an accretion disk in the equatorial plane of the primary, these intersections will occur at different times and radii for an inclined eccentric trajectory with ``beat'' patterns in the time domain. For example, taking a moderate eccentricity $e\sim0.3$ and the orbital period and the MBH mass as 1 hour and $10^6 M_\odot$, respectively, the pericenter precession will complete a cycle in roughly 44 hours, so the beat pattern will span $\sim 88$ eruptions. An example of such an orbit and its intersection times and radii are plotted in Figs~\ref{fig:orbit} and \ref{fig:interstime}.

Nevertheless, the EM flares are not necessarily produced immediately after the intersection, so identifying this secondary frequency may be subtle, as already discussed in Section \ref{sssec:uncert}. In particular, since the eccentric orbits intersect the disk at different radii and thus interact with a plasma slab with different temperatures, densities, and geometrical thickness, the \textit{delays between impact and radiation release will vary among impacts}. This introduces additional timing features into the model that depend on the disk properties, as exemplified in the case of the fit of the parameters of the putative eccentric MBH binary in OJ 287 \citep{dey2018authenticating,2023MNRAS.526.2754Z}.

Another non-trivial timing feature may emerge from the fact that the orbital plane of the perturber will generally precess due to the spin-orbital coupling. This is to leading post-Newtonian order and mass ratio given by the Lense-Thirring frequency \citep{1988NCimB.101..127D}
\begin{align}
\begin{split}
    & \Omega_{\rm LT} = \frac{8 \pi^2 G M a}{P_{\rm orb}^2 c^3 (1- e^2)^{3/2}} 
    \\& \phantom{\Omega_{\rm LT} } 
    \sim \frac{0.11}{\rm hour} \frac{a}{(1 - e^2)^{3/2}}  \left(\frac{M}{10^6\,M_{\odot}} \right) \left(\frac{P_{\rm orb}}{1\,\text{hour}} \right)^{-2} \,,
\end{split}
\end{align}
where $a \in (-1,1)$ is the dimensionless spin of the MBH.
For an accretion disk aligned with the MBH spin this does not lead to significant effects in the temporal spacing of the flares. However, if the accretion disk is not aligned with the equatorial plane, the interplay of the differential precession of the accretion disk \citep{1975ApJ...195L..65B} and the precession of the orbit will lead to additional intricate variability, a fact that was used by \citet{Franchini_2023} to explain some of the temporal properties of the QPEs in GSN 069, eRO-QPE1, eRO-QPE2, and RX J1301.9+2747.

\section{Conclusion}\label{sec:conclusion}
EMRI observations through the LISA GW detector will provide a unique probe for fundamental physics in our Universe. Environment-rich formation channels of EMRIs would potentially give rise to their EM counterparts. These may be observable as soft X-ray ($\sim 10^{-1}$ keV) band, low-frequency phenomena known as QPEs, and/or the less-abrupt QPOs, which emit semi-regular X-ray signals from both quiescent and active galactic nuclei. In this paper, we estimated the frequency band, in which a population of EMRIs in the Universe, as described in the \texttt{M1} EMRI catalog of \cite{Babak_2017}, will most likely emit a QPE or QPO signal in the soft X-ray band ``today'' (in 2024). After establishing this frequency band to be $\approx 0.46 \pm 0.22$ mHz, we did a converse analysis on the well-known QPO source, RE J1034+396, treating it as an EMRI to estimate its CO's mass to be $\approx 46 M_\odot$, and calculated a prior-predictive distribution on its LISA-band SNR with a best-fit value of $\rho \approx 14$, highlighting it as a potential multi-messenger EMRI source. Given the estimate on the frequency band, we explored the possibility of observing LISA-EMRI relevant QPEs and QPOs in the EM spectrum through a variety of current and upcoming detectors in the X-ray, UV, and even radio bands. The impact of several EMRI parameters currently unaccounted for in our estimates, such as smaller masses of the CO, and the effect of eccentricity and inclination on the QPE/QPO flares, were also discussed. Our study thus attempted to strengthen the connection between EMRIs and QPEs/QPOs, highlighting them as potential multimessenger inference targets. If such observations become feasible in the future, it would greatly enhance the science output of the recently adopted LISA observatory.

\section*{Acknowledgements}

SK acknowledges the computational resources made accessible by the National University of Singapore's IT Research Computing group and the support of the NUS Research Scholarship (NUSRS). MZ acknowledges the financial support of the GA\v{C}R Junior Star grant no. GM24-10599M. Fig. \ref{fig:orbit} was generated with the help of the {\texttt{KerrGeodesics}} package within the BHPToolkit (bhptoolkit.org). VW and SK are grateful to the organizers for their invitation to the \textit{1st Trieste meeting on the physics of gravitational waves} where the idea of this paper was conceived. 

\section*{Data Availability}

The results of this study are fully reproducible and are made available on a Github repository called \href{https://github.com/perturber/EMRI_EM_Counterparts}{\texttt{EMRI\_EM\_Counterparts}}\footnote{\href{https://github.com/perturber/EMRI_EM_Counterparts}{https://github.com/perturber/EMRI\_EM\_Counterparts}}. \cite{Babak_2017}'s \texttt{M1} EMRI catalog as modified by \cite{pozzoli2023computation} can be obtained by contacting the authors\footnote{Federico Pozzoli: \href{mailto:fpozzoli@uninsubria.it}{fpozzoli@uninsubria.it}}, but we have also made available the back evolution trajectories data for our analysis in Section~\ref{sec:backevolve} on \href{https://zenodo.org/records/10812229?token=eyJhbGciOiJIUzUxMiJ9.eyJpZCI6IjllOTQ1ZDM5LTRmYTktNDIzMy04ZTUwLTlhODc4ODMzNDBhNyIsImRhdGEiOnt9LCJyYW5kb20iOiIwMjk4NWUyYWQzNmUzNzJlOWQzZTA0ZTAxNmExYjdjNiJ9.85VYgKGZ-1CluoL1X5fsThPYyrRJ1ylX_GIgTwr8T18ooEsDK3MbXH-BeLrpsETnzqLaoTUQgDj_jN2ddVgK7A}{Zenodo}\footnote{\href{https://doi.org/10.5281/zenodo.10812229}{doi.org/10.5281/zenodo.10812229}}.



\bibliographystyle{mnras}
\bibliography{bibliography} 



\appendix

\section{Additional priors on RE J1034+396 parameters}\label{app:REJ}
\begin{figure*}
\centering
\includegraphics[width=1.0\textwidth]{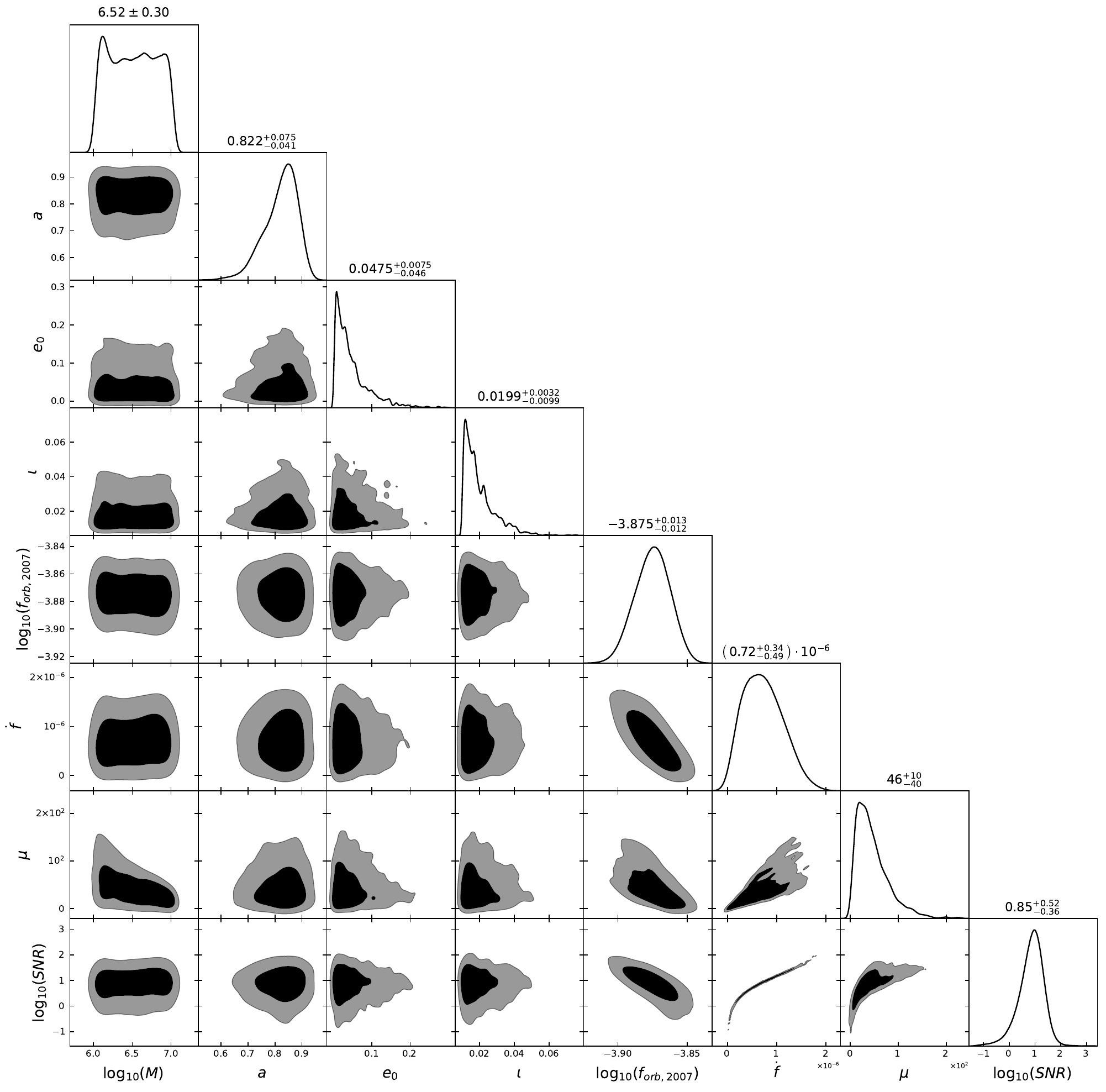}
\caption{Distributions of $\{M,a,e_0,\iota,f_{\rm orb,2007},\dot{f}_{\rm orb,obs},\mu,\text{SNR}\}$ of RE J1034+396 modeled as an extreme-mass-ratio inspiral. $\{M,f_{\rm orb,2007},\dot{f}_{\rm orb,obs}\} $ are estimated from current observational constraints on RE J1034+396 and distributions of $\{a,e_0,\iota\}$ are based on expected properties of EMRIs as described in the text. The thus obtained prior-predictive distribution on $\mu$ is described in the seventh panel, and the corresponding LISA-band SNR in the final panel.
}
\label{fig:SNRdist}
\end{figure*}

Here, for the QPO source RE J1034+396, we determine the priors on the MBH dimensionless spin parameter, $a$, and the eccentricity and inclination of the CO's orbit, i.e. $e_0$ and $\iota$ respectively. We use a right skewed-normal prior on $a$ with a mean of 0.9. This is based on evidence suggesting that MBHs in AGNs are generally expected to be endowed with a large spin \citep[see, for example, ][]{Volonteri_2005, Reynolds_2013, Brenneman_2013}. Our model assumes prograde orbits, i.e. orbits in the same direction as the MBH spin. This is expected for EMRI sources, in which CO is either captured by the accretion disk surrounding the MBH or forms \textit{in-situ} within the disk and migrates towards tighter orbits (see references for formation channels (iii)-(iv) as described in Section \ref{sec:introduction}, and \citeauthor{2016ApJ...819L..17B}, \citeyear{2016ApJ...819L..17B}). For the eccentricity and inclination of the CO's orbit, we choose exponentially decaying distributions with means 0.05 and 0.01 for eccentricity and inclination respectively, allowing for only slightly inclined and eccentric orbits. This builds on the evidence that small objects orbiting a massive body in the presence of a disk lose their energy and angular momentum through interactions with the disk, circularizing their orbits and decreasing their inclination (\citeauthor{2004ApJ...602..388T}, \citeyear{2004ApJ...602..388T}; see also \citeauthor{Kley_2012}, \citeyear{Kley_2012} for a review). Energy and angular momentum loss through GW radiation in EMRIs further assists this circularization, such that environment-rich EMRIs can be expected to have small eccentricities and inclinations~\citep{Pan_2021}. As noted in Section \ref{sec:REJ1034396}, the correlations of the estimated parameters ($\mu, \rho$) with the set $\{a,e_0,\iota\}$ is small $(|C_{ij}| \lesssim 0.02)$, and thus the choice of prior does not change our results significantly. 
The full 8-dimensional prior-predictive estimates on the RE J1034+396 parameter set $\{M, a, e_0, \iota, f_{\rm orb, 2007}, \dot{f}, \mu, \rm{SNR}\}$ are provided in Fig. \ref{fig:SNRdist}.

\bsp	
\label{lastpage}
\end{document}